\documentclass[prb,twocolumn,superscriptaddress, preprintnumbers,letterpaper,twoside]{revtex4}

\usepackage{times}
\usepackage{amsmath}
\usepackage{amsthm}
\usepackage{amssymb}
\usepackage{amsbsy}
\usepackage{amsfonts}
\usepackage{bm}
\usepackage{graphicx}
\usepackage{multirow}
\usepackage{verbatim}
\usepackage{hyperref}
\hypersetup{
        colorlinks=true,
}
\usepackage{tikz}

\renewcommand{\vec}[1]{{\mathbf #1}}

\newcommand{\comments}[1]{}

\newcommand{\GL}{\mathrm{GL}}

\begin{document}


\title{Chirality-Protected Majorana Zero Modes at the Gapless Edge of 
Abelian Quantum Hall States}

\author{Jennifer Cano}
\affiliation{Department of Physics, University of California, Santa Barbara,
California 93106, USA}
\author{Meng Cheng}
\affiliation{Microsoft Research, Station Q, Elings Hall,
University of California, Santa Barbara, California 93106-6105, USA}
\author{Maissam Barkeshli}
\affiliation{Microsoft Research, Station Q, Elings Hall,
University of California, Santa Barbara, California 93106-6105, USA}
\author{David J. Clarke}
\affiliation{Condensed Matter Theory Center, Department of Physics, University of Maryland, College Park}
\affiliation{Joint Quantum Institute, University of Maryland, College Park, Maryland 20742, USA}
\affiliation{Microsoft Research, Station Q, Elings Hall,
University of California, Santa Barbara, California 93106-6105, USA}
\author{Chetan Nayak}
\affiliation{Microsoft Research, Station Q, Elings Hall,
University of California, Santa Barbara, California 93106-6105, USA}
\affiliation{Department of Physics, University of California, Santa Barbara, California 93106, USA}

\begin{abstract}
We show that the $\nu=8$ integer quantum Hall state can support Majorana zero modes at domain walls between its two different stable chiral edge phases without superconductivity.  This is due to the existence of an edge phase that does not support gapless fermionic excitations; all gapless excitations are bosonic in this edge phase.  Majorana fermion zero modes occur at a domain wall between this edge phase and the more conventional one that does support gapless fermions.  Remarkably, due to the chirality of the system, the topological degeneracy of these zero modes has exponential protection, as a function of the relevant length scales, in spite of the presence of gapless excitations, including gapless fermions.  These results are compatible with charge conservation, but do not require it.  We discuss generalizations to other integer and fractional quantum Hall states, and classify possible mechanisms for appearance of Majorana zero modes at domain walls.
\end{abstract}

\maketitle


\section{Introduction.}

Majorana zero modes have been the focus of recent theoretical and experimental efforts
\cite{Alicea12a,Beenakker13a}, motivated in part by their potential applications to topological
quantum information processing \cite{Kitaev97,Freedman98,Nayak08}.
A Majorana operator $\gamma$ is a self-hermitian fermionic operator,
$\gamma=\gamma^\dagger \,\, , \,\,\,\, \gamma^2=1$.
It is a zero mode if it commutes with the Hamiltonian $H$, i.e. if
$[H,\gamma]=0$. This occurs most naturally in superconductors,
where the eigenstates are superpositions of particles and holes
so that $\gamma = c + c^\dagger$ where $c$ is the annihilation operator for electrons. For this reason, experimental attention
\cite{Jang11,Mourik12,Rokhinson12,Deng12,Churchill13,Das12,Finck12} has been focussed on
putative $p+ip$ superconductors \cite{DasSarma06a} and on proximity-induced superconductivity
in topological insulator surface states \cite{Fu08}, semiconductor
quantum wells \cite{Sau10a}, and semiconductor nanowires \cite{Lutchyn10,Oreg10}.
The superconductivity need not be long-range-ordered; Majorana zero modes can also
occur in systems that only have quasi-long-range-ordered
superconductivity \cite{Fidkowski11b,Sau11}.
In fact, superconductivity is not necessary at all: Ising topological order
supports an analog of superconductivity \cite{Moore91,Greiter92} that is sufficient to support Majorana zero modes.
This includes the Moore-Read state \cite{Moore91,Nayak96c,Read96,Bonderson11a} and the
anti-Pfaffian state \cite{LeeSS07,Levin07}, which are candidate descriptions of the
$\nu=5/2$ fractional quantum Hall state. Certain extrinsic defects in bilayer Abelian fractional quantum 
Hall systems\cite{Barkeshli12,Barkeshli13a} or $\mathbb{Z}_2$ toric code models\cite{bombin2010} can also localize 
Majorana zero modes without any superconductivity.

Recently, it was shown that an Abelian topological phase, such as an integer quantum Hall state,
can have multiple stable chiral edge phases \cite{Cano13b}. By tuning parameters at the edge, one can
drive the system through edge phase transitions that leave the bulk unaffected.
One of the simplest examples of this is the $\nu=8$ integer quantum Hall state, which has an
edge phase, which we will call the $I_8$ phase, that is continuously connected to the edge of a
non-interacting electron system, and a second edge phase, the $E_8$ phase,
which has only bosonic excitations. This raises the question, then, of what happens when there
is a domain wall at the edge, with the $I_8$ phase on one side and $E_8$ on the other,
as depicted in Fig. \ref{fig:Edge-Sections}. If the $I_8$ phase lies upstream, then a low-energy
fermionic excitation cannot propagate through the domain wall to the $E_8$ side
since all fermionic excitations are gapped in the latter region. The $I_8$ regions have a fermion
parity that is conserved by the dynamics of the edge, so long as no electrons tunnel in
from external leads or localized bulk states. Moreover, the ground state of each fermion parity
has the same energy since there are no gapless excitations in either phase that can measure
the fermion parity -- only $hc/2e$ vortices can do that. Thus, each such region has the same Hilbert space
as a pair of Majorana zero modes at the ends of a superconducting nanowire. Surprisingly, we find that even in the presence of gapless (fermionic!) degrees of freedom, the topological degeneracy is protected exponentially: the energy splitting of the ground states decays exponentially with the separations between the domain walls. The topological protection of Majorana zero modes in this scenario is attributed to the chirality of the edge modes, which require a 2D bulk phase to exist. This should be compared with the scenario considered in Refs. [\onlinecite{Fidkowski11b,Sau11}], where Majorana zero modes occur in number-conserving one-dimensional wires coupled to quasi-long-ranged superconductors, but are protected only algebraically.

As shown in Ref. [\onlinecite{Cano13b}], there are also fermionic fractional quantum Hall states that admit edge phases without gapless fermionic excitations. These also support Majorana zero modes at domain walls between
bosonic and fermionic edge phases. Finally, there are fractional quantum Hall states with edge phases
in which only a subset of the bulk quasiparticle types are gapless (the subset that braids trivially
with a non-trivial bosonic quasiparticle that condenses on the edge); these, too, have a topological degeneracy associated with domain walls.

The paper is organized as follows: we begin by reviewing the edge theory of quantum Hall states and the two edge phases ($E_8$ and $I_8$) of the $\nu=8$ integer quantum Hall state in Sec. \ref{sec:edgephases}. In Sec. \ref{sec:degeneracy}
 we establish that there are Majorana zero modes at the domain walls between $E_8$ and $I_8$ edge phases by directly solving a representative model of the edge phases in the low-energy limit.  Sec. \ref{sec:splittings}
 and Sec. \ref{sec:other-gap-terms} address the stability of the topological degeneracy against perturbations.  In Sec. \ref{sec:fqh} and \ref{sec:domain-walls-same-qps} we discuss generalizations to other Abelian fractional quantum Hall states. In \ref{sec:classification}, we present a synthesis of the topological degeneracy of domain walls on the edge theory.


\section{Edge Phases of the $\nu=8$ integer quantum Hall state}
\label{sec:edgephases}

We begin by recalling some facts about the edge phases
of the $\nu=8$ integer quantum Hall state \cite{Cano13b}.
Low-energy edge excitations of an Abelian quantum Hall state
may be described by the chiral Luttinger liquid
effective action:
\begin{multline}
\label{eqn:bosonic-general}
S_{LL} = \int dx\, dt \biggl(\frac{1}{4\pi}K_{IJ} \partial_t \phi^I \partial_x \phi^J -
\frac{1}{4\pi}V_{IJ}\partial_x \phi^I \partial_x \phi^J\\ + \frac{1}{2\pi} t_I \epsilon_{\mu\nu} \partial_\mu \phi^I A_\nu
\biggr).
\end{multline}
The fields in this action satisfy the periodicity condition $\phi^I \equiv \phi^I + 2\pi n^I$
for $n^I \in \mathbb{Z}$ and the equal-time commutation relation $[\phi_I(x), \partial_y\phi_J(y) ] = 2\pi i K^{-1}_{IJ}\delta(x-y)$.
An edge phase is characterized by an equivalence class of a
positive-definite symmetric integer \emph{K-matrix}, and integer \emph{charge vector} $t$,
with respect to $\mathrm{GL}(N,\mathbb{Z})$ basis transformations $\tilde{K}=W^T K W, \tilde{t}=W^Tt$, 
with $W\in \mathrm{GL}(N,\mathbb{Z})$. Such transformations are induced by invertible
changes of variables $\phi^I=W^I_{\ J}{\tilde \phi}^J$ that preserve
the periodicity of the fields $\phi^I$.
The charge vector $t$ determines the coupling to the external electromagnetic field, and the velocity matrix $V_{IJ}$ is a real matrix that determines the velocities of the edge modes
and, when the theory is not fully chiral,
also determines the scaling dimensions of operators.

It is useful to characterize these phases by lattices $\Lambda$ rather than equivalence classes
of K-matrices. Let $e_{I}^a$ be the eigenvector of $K$ corresponding to eigenvalue
$\lambda_a$: $K_{IJ} e_{J}^a = \lambda^a e_{I}^a$. We normalize $e_{J}^a$ so that
$e_{J}^a e_{J}^b = \delta^{ab}$ and define a metric $g_{ab} = \lambda_a \delta_{ab}$.
Then, $K_{IJ} =  g_{ab} e_{I}^a e_{J}^b$ or, using vector notation, $K_{IJ} =  {\bf e}_I \cdot {\bf e}_J$.
The metric $g_{a b}$ defines a bilinear form on the lattice $\Lambda$ --
this just means we can multiply two lattice vectors $\vec{e}_I, \vec{e}_J$ together using the metric, $\vec{e}_I \cdot \vec{e}_J = e_I^a g_{a b} e_J^b$.
The $N$ vectors ${\bf e}_I$ define a lattice $\Lambda = \{ m_I {\bf e}_I  | {m_I} \in \mathbb{Z}\}$.
The $\GL(N,\mathbb{Z})$ transformations $K \rightarrow W^T K W$ are simply basis changes
of this lattice, so we can equally well describe edge phases by equivalence classes of $K$-matrices
or by lattices $\Lambda$. 

The connection of quantum Hall edge phases to lattices
can be exploited more easily if we make the following change of variables,
$X^a = e_{I}^a \phi^I$, in terms of which the action takes the form
\begin{equation}
\label{eqn:bosonic-X1}
S = \frac{1}{4\pi} \int dx\,dt \biggl(g_{ab} \partial_t X^a \partial_x X^b -
v_{ab}\partial_x X^a \partial_x X^b.
\biggr)
\end{equation}
The variables $X^a$ satisfy the periodicity condition ${\bf X} \equiv {\bf X} + 2\pi {\bf y}$
for ${\bf y} \in \Lambda$ and $v_{ab}\equiv V_{IJ} f^I_{a} f^J_{b}$,
where $f_{a}^I$ are basis vectors for the dual lattice $\Lambda^*$, satisfying
$f^I_{a} e_J^a = e_{L a} (K^{-1})^{L I} e_J^a = \delta^I_J$.

We now focus on the $\nu=8$ integer quantum Hall state. There are two possible choices for $(K_{IJ},{t_I})$ that are consistent with the same bulk phase~\cite{Cano13b}. The first is $K_{IJ}=\delta_{IJ}$ and ${t_I}=1$, which is continuously
connected to the edge of the $\nu=8$ state of non-interacting electrons.
We will call this the $I_8$ phase. The corresponding lattice is just the $8$-dimensional
hypercubic lattice $\mathbb{Z}^8$. For later convenience,
we make the basis change $W=\text{diag}(-{\bf 1}_{3},{\bf 1}_{5})$, which leaves
$K_{IJ}=\delta_{IJ}$ unchanged but transforms $t_I$ to $(-1,-1,-1,1,1,1,1,1)$.
The second phase has $K_{IJ}=K^{E_8}_{IJ}$, where
$K^{E_8}$ is the Cartan matrix of $E_8$, given explicitly in Appendix~\ref{sec:matrices}. The corresponding charge vector is $t=(4, -2, 0, 0, 0, 0, 0, 0)$. The lattice is the root lattice of the $E_8$ Lie algebra, hence the name.

The edge phase transition between these phases occurs when an additional non-chiral pair of modes comes
down in energy and interacts with the $8$ chiral modes.
Such modes are normally present but, in general, some non-chiral combination of right- and left-moving modes will be gapped at low energies. The particular combination that gets gapped
determines the phase of the remaining $8$ gapless chiral modes. 
For the sake of concreteness, let us begin in the ${E_8}$ edge phase:
\begin{multline}
\label{eqn:S_0}
{S_0} = \int dx\,dt \biggl(\frac{1}{4\pi}\left( K^{E_8} \oplus \sigma_z\right)^{}_{IJ} \partial_t \phi^I \partial_x \phi^J \\ -
\frac{1}{4\pi}V_{IJ}\partial_x \phi^I \partial_x \phi^J 
+ \frac{1}{2\pi}{t_I}\epsilon_{\mu\nu} \partial_\mu \phi^I A_\nu
\biggr).
\end{multline}
Now, $I,J =1, \ldots, 10$, and $\sigma_z = \text{diag}(1,-1)$ is the $K$-matrix for the
non-chiral pair of modes. We assume that $t_{9}=t_{10}=-1$. (This corresponds to
adding a non-chiral pair with $\tilde{t}_{9}=\tilde{t}_{10}=1$ to the $I_8$ state.)
We now consider perturbations that could gap a non-chiral combination of modes.
We focus on $S={S_0}+{S_\text{g}}$, where
\begin{multline}
\label{eqn:two-cosines}
S_\text{g} =  \int\! dx\,dt\,  \big[{u_E}\cos(\phi_9 + \phi_{10})\\ +
{u_I}  \cos(-{\phi_1}+ \phi_9 + 3\phi_{10})\bigr]
\end{multline}
If ${u_E}\gg {u_I}$ or is more relevant (which
is determined by the matrix $V_{IJ}$), then $\phi_9$ and $\phi_{10}$ will be gapped and
the system will be in the $E_8$ phase. On the other hand, if ${u_I}\gg {u_E}$ or is more relevant,
then the system will be in the $I_8$ phase \cite{Cano13b}, which may be seen as follows.
We first note that ${I_8}\oplus \sigma_z = {(W^8)^T}(K^{E_8} \oplus \sigma_z)W^8$, where
the explicit form of $W^8$ is given in Appendix~\ref{sec:matrices}. If we make the change of variables
$\phi_I = W^8_{IJ} \tilde{\phi}_J$, then Eq. (\ref{eqn:S_0}) takes the form
\begin{multline}
\label{eqn:S_0-tilde}
{S_0} = \int dx\,dt \biggl(\frac{1}{4\pi}\left( I_8 \oplus \sigma_z\right)^{}_{IJ} \partial_t \tilde{\phi}^I \partial_x \tilde{\phi}^J \\ -
\frac{1}{4\pi}\tilde{V}_{IJ}\partial_x \tilde{\phi}^I \partial_x \tilde{\phi}^J 
+ \frac{1}{2\pi}\tilde{t}_{I}\epsilon_{\mu\nu} \partial_\mu \tilde{\phi}^I A_\nu
\biggr),
\end{multline}
where $\tilde{t}\equiv (W^8)^Tt = \text{diag}(-{\bf 1}_{3},{\bf 1}_{5})$
and $\tilde{V}_{IJ} \equiv (W^8)^T_{IK} V_{KL} W^8_{LJ}$.
Then Eq.~(\ref{eqn:two-cosines}) takes the form:
\begin{multline}
\label{eqn:two-cosines-tilde}
S_\text{g} =  \int\! dx\,dt\,  \big[{u_E}\cos(\tilde{\phi}_{1} + \tilde{\phi}_{2} \ldots + \tilde{\phi}_{9}
+ 3 \tilde{\phi}_{10})\\ +
{u_I}  \cos(\tilde{\phi}_{9} + \tilde{\phi}_{10})\bigr]
\end{multline}
Thus, $u_I$ gives a gap to $\tilde{\phi}_{9}, \tilde{\phi}_{10}$, leaving behind the theory with
$K=I_8$. The general criterion for determining whether the low-energy theory is described by $K=I_8$ or $K^{E_8}$, given a particular gapping term, is given in Appendix~\ref{sec:find-Keff}.


\section{Ground state degeneracy of an edge with multiple domain walls}
\label{sec:degeneracy}

\begin{figure}[t]
  \includegraphics[width=8cm]{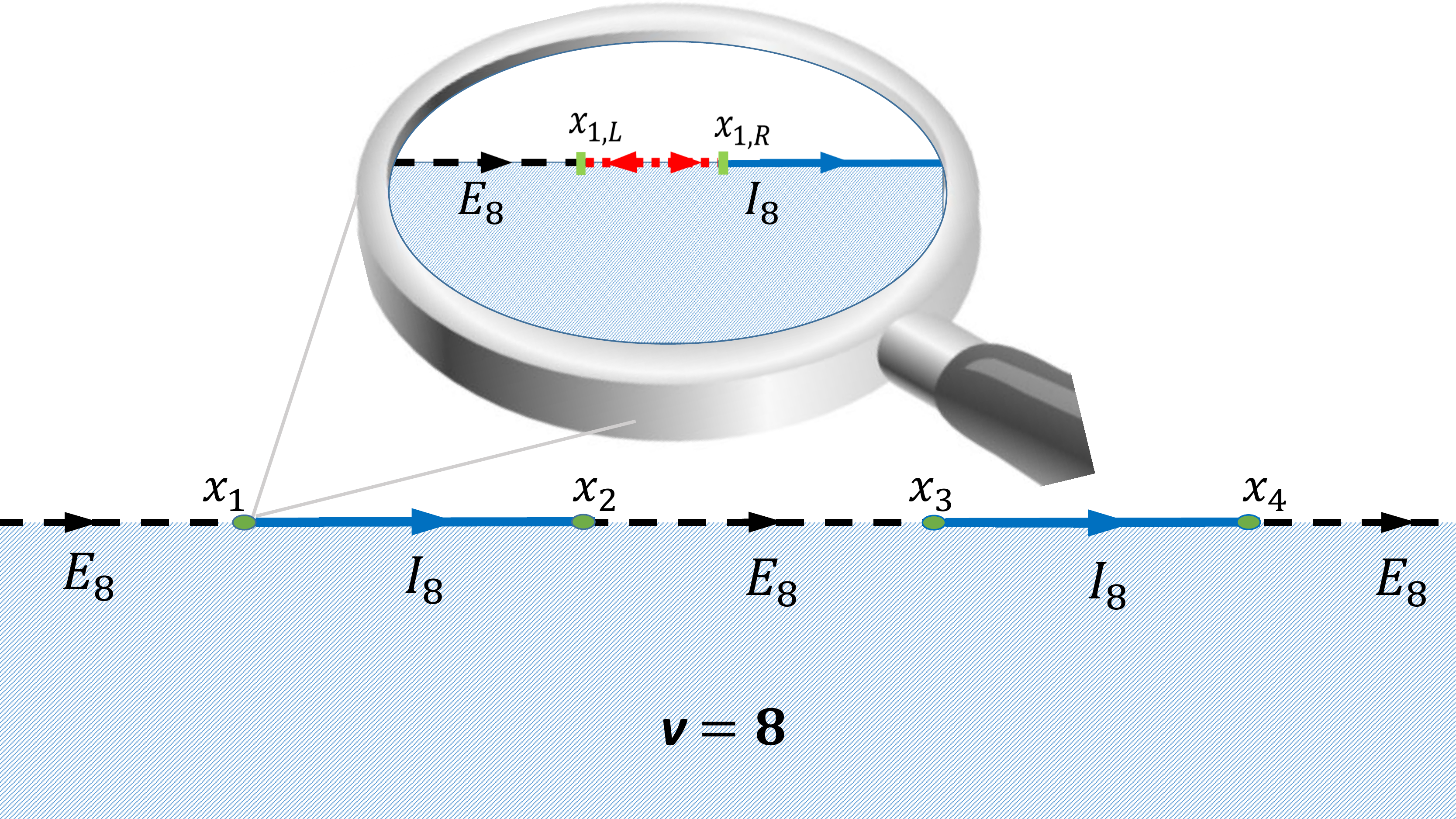}
  \caption{The edge of a conventional $\nu=8$ integer quantum Hall state is divided into alternating sections
in the two different stable chiral edge phases: the $I_8$ and $E_8$ phases. The domain wall
centered at ${x_i}$ is a small interval from $x_{i,L}$ to $x_{i,R}$ in which there is a counter-propagating mode,
due to edge reconstruction. As discussed in the text, this configuration has degenerate ground states,
which can be interpreted as Majorana zero modes at the domain walls.}
  \label{fig:Edge-Sections}
\end{figure}

We now consider a quantum Hall droplet whose edge is divided into multiple sections,
with domain walls at $x_1$, $x_2$, $\ldots$, as depicted in Fig. \ref{fig:Edge-Sections}.
For later convenience, we will take the domain
walls to have width $2a$, extending from $x_{i,L}$ to $x_{i,R}$, with $x_{i,R} - x_{i}=x_{i} - x_{i,L}=a$.
The regions $C_i$ between the domain walls lie between $x_{i,R}$ and $x_{i+1,L}$.
We will assume that $C_{I_8} \equiv \bigcup_n C_{2n-1}$ is in the $I_8$ phase while
$C_{E_8}\equiv \bigcup_n C_{2n}$ is in the $E_8$ phase. To arrange this, we take an effective action of the form  $S={S_0}+{S_\text{g}}$ with
${u_E}, {u_I} \rightarrow {u_E}(x), {u_I}(x)$, where
${u_I}(x) = u\chi_\text{o}(x)$, ${u_E}(x) = u\chi_\text{e}(x)$,
and $\chi_\text{o}(x)=1$ for $x\in C_{I_8}$ and $\chi_\text{o}(x)=0$ otherwise
while $\chi_\text{e}(x)=1$ for $x\in C_{E_8}$ and $\chi_\text{e}(x)=0$ otherwise.
The parameter $u$ is assumed to be large so that the fields are pinned to the minima
of the corresponding cosines everywhere except at the domain walls, $x_{i,L}<x<x_{i,R}$.

Let us first consider one domain wall between $C_1$ and $C_2$. Since the arguments of the cosines in $C_1$ and $C_2$ do not commute, only one cosine can be pinned in 
a single one of its minima, while the other must be in a superposition of its minima. Suppose that we choose a
basis in which ${\phi_9}+\phi_{10}$ in $C_1$ is pinned
to one of the minima of the cosine, so that ${\phi_9}+\phi_{10} = 2\pi n$, for some
$n\in \mathbb{Z}$. Then $-{\phi_1}+{\phi_9}+3\phi_{10}$ in $C_2$ will be in a superposition of minima connected by
the action of the operator $\cos({\phi_9}+\phi_{10})$, i.e. shifted by $4\pi$.
To be more precise, note that the sectors of the theory
can be labeled $|\tilde{p}_{1},\ldots,\tilde{p}_{8},-{\phi^0_1}+{\phi^0_9}+3\phi^{0}_{10}\rangle$,
where $\tilde{p}_{I}$ is the constant part (i.e. the zero mode) of ${\partial_x}\tilde{\phi}_I$ (physically, it is the charge density in mode $I$) and $-{\phi^0_1}+{\phi^0_9}+3\phi^{0}_{10}$
is the constant part of the corresponding combination of fields.
In this basis, the ground states are ${\sum_m}|{\bf 0}_{8}, 4\pi m\rangle$ and
${\sum_m}|{\bf 0}_{8}, 2\pi + 4\pi m\rangle$. Here ${\bf 0}_8$ denotes collectively the values of $\tilde{p}_1, \dots, \tilde{p}_8$. Either of these states
will satisfy both cosines. Thus, we conclude that there is a two-fold ground state degeneracy
in the $u\rightarrow\infty$ limit. To generalize this analysis to arbitrary number of domain walls, it is important to take into account the global structure of the moduli space of the bosonic fields~\cite{Fidkowski11b}. Below we will use a different, perhaps more physical, method to count the ground-state degeneracy. 

First, we give a physical interpretation that explains why this degeneracy is robust.
Note that, in the basis of boson fields that we have chosen, the fields ${\phi_1},\ldots,{\phi_8}$
can only create bosonic excitations. Fermionic excitations necessarily involve $\phi_9$ and $\phi_{10}$,
so there are no fermionic excitations in $C_{E_8}$ in the ${u_E}\rightarrow\infty$ limit.
Meanwhile, the coupling $u_I$ preserves the fermion parity of $\phi_9$ and $\phi_{10}$
in $C_{I_8}$. It tunnels a bosonic excitation from $\phi_9$ and $\phi_{10}$ to
${\phi_1},\ldots,{\phi_8}$ and can, therefore, be viewed as analogous to
the pair tunneling term that couples a semiconductor wire to a superconducting wire \cite{Fidkowski11b}.
Since the fermion parity of $\phi_9$ and $\phi_{10}$ in $C_{I_8}$ cannot be changed
by tunneling into $C_{E_8}$ or by tunneling to ${\phi_1},\ldots,{\phi_8}$,
it is conserved.  
Thus, if there are $2k$ domain walls, the total ground state degeneracy is $2^{k-1}$: for fixed total fermion parity, each of the $k$ regions in $C_{I_8}$ can have even or odd parity. Note that states with different total fermion parity are not expected to be degenerate; we examine this point in detail in Sec~\ref{sec:total-droplet}.

As we discuss in the next paragraph,
there are no terms that can be added to the Hamiltonian that
would violate this low-energy conservation law without closing the energy gap to the counter-propagating modes in $C_{E_8}$. Moreover, as we will see in the next section, phase slips in the $E_8$ phase do not cause any splitting due to chirality.  In summary, there
are no local terms that can be added to the Hamiltonian that would cause an energy splitting
between the even and odd parity ground states. Therefore, the degeneracy is robust. 
This can be recast in more formal terms by introducing the operators
\begin{eqnarray}
\label{eqn:degeneracy-ops}
A_j &=& \exp\left[{\frac{i}{2}\int_{x_{2j-1,L}}^{x_{2j,R}} {\partial_x}(\phi_{9}+\phi_{10})}\right]\, , \cr
B_j &=& \exp\left[{\frac{i}{2}\int_{x_{2j,L}}^{x_{2j+1,R}} {\partial_x}({\tilde \phi}_{9}+{\tilde \phi}_{10})}\right]\\
 \nonumber &=& \exp\left[{\frac{i}{2}\int_{x_{2j,L}}^{x_{2j+1,R}} {\partial_x}(-{\phi_1} + {\phi_9} + 3\phi_{10})}\right]
\end{eqnarray}
$A_j$ operators just measure the total fermion parity of the corresponding $I_8$ region, while $B_j$ measures the fermion parity stored in the $\tilde{\phi}_9$ and $\tilde{\phi}_{10}$ modes in the corresponding $E_8$
region and effectively tunnels a fermion across this $E_8$ region. These operators satisfy 
\begin{equation}
	\begin{gathered}
		{A_j^2} = {B_j^2}=1,\\ 
		\{{A_j},{B_j}\}=\{A_{j},B_{j-1}\}=0\\
		[{A_j},{B_k}]=0 \text{ for } k\neq j, j-1\\
	\end{gathered}
		\label{}
\end{equation}
Furthermore $[H,A_{j}]=[H,B_{j}]=0$, so these operators form an algebra over the ground state subspace. Therefore, the ground state degeneracy is $2^{k-1}$ if there are $2k$ domain walls. We note that since the degeneracy is $2^{k-1}$, it follows that if there is only one $I_8$ region and one $E_8$ region, there is no degeneracy.
While the two possible fermion parities of an $I_8$ region are degenerate (assuming that there are other $I_8$ regions),
there is no degeneracy between different electron numbers of the entire droplet. (The droplet has a fixed electron
number, not merely a parity, since charge is conserved.) We discuss the splitting of this degeneracy in Section
\ref{sec:total-droplet}.

Following Ref.~\onlinecite{Clarke13a}, we can define Majorana fermion operators.
When the coefficients of the cosines $u_I$, $u_E$ are large,
their arguments are $2\pi$ multiplied by integer-valued operators  $\hat{m}_{i}$, $\hat{n}_{i}$:
\begin{equation}
\label{eqn:cos-values}
2\pi \hat{m}_{i} =({\tilde \phi}_{9}+{\tilde \phi}_{10})_{C_{2i-1}}\, , \,\,
2\pi \hat{n}_{i} = ({\phi_9} + \phi_{10})_{C_{2i}}
\end{equation}
At the domain wall between $C_0$ and $C_{1}$, we can define the Majorana fermion operator:
\begin{equation}
	\gamma_{1} \equiv e^{i\pi(\hat{m}_{1} -\hat{n}_{0})}.
\end{equation}
Defining $\gamma_2$, $\gamma_3$, $\ldots$ similarly in terms of the arguments of the cosines that flank
the corresponding domain walls, we see that the operators introduced in the previous paragraph can be expressed as:
${A_1} =i{\gamma_1} {\gamma_2}$, ${B_2} =i{\gamma_2} {\gamma_3}$.

If we express $\gamma_1$ in terms of the original electron operators, however, we see that:
\begin{multline}
\label{eqn:MZM-def}
	\gamma_1 = e^{-\frac{i}{2} {\phi_1}(x_{1,R})}\,\times \,e^{i\phi_{10}(x_{1,R})}\,\times\\
	\,e^{\frac{i}{2}[\phi_{9}(x_{1,R}) - \phi_{9}(x_{1,L})]} \,
	\,e^{\frac{i}{2}[\phi_{10}(x_{1,R}) - \phi_{10}(x_{1,L})]}
\end{multline}
From the first factor in this expression, it is apparent that the Majorana fermion operator
is not local in terms of the original electron operators.
In a system coupled to a $3$D superconductor, we would have a similar expression,
but with the second term on the right-hand-side replaced by $e^{i\theta/2}$, where $\theta$ is the phase of the superconducting order parameter,
which can be treated as a classical number. However, in the present case,
the fluctuating non-local expression $e^{-\frac{i}{2} {\phi_1}(x_{1,R})}$ is necessary to
relate the Majorana fermion $\gamma_1$ to the electron operator
at the domain wall, $e^{i\phi_{10}(x_{1,R})}$ (together with
the local fluctuation $e^{\frac{i}{2}[\phi_{9}(x_{1,R}) - \phi_{9}(x_{1,L})]} \,
\,e^{\frac{i}{2}[\phi_{10}(x_{1,R}) - \phi_{10}(x_{1,L})]}$).

One might worry that the ground state degeneracy is unstable to arbitrary perturbations, since there are gapless fermionic excitations in the $I_8$ regions. We show in Sec~\ref{sec:splittings} that this is not the case and show, moreover,
that the finite-size splitting between nearly-degenerate ground states decays exponentially with size. 

\section{Scattering problem at the domain wall}
\label{sec:scattering}

Now that we have shown explicitly the topological degeneracy of domain walls between $E_8$ and $I_8$ edge phases, we return to the heuristic observation made in the introduction: Intuitively, Majorana zero modes must exist at $I_8$-$E_8$ domain walls in order to absorb low-energy fermionic excitations originating within the $I_8$ regions.
It is instructive to see how this occurs by solving the scattering problem of the fields at the domain wall, and deriving an ``$S$ matrix''. 

We focus on the behavior of the fields in the vicinity of the domain wall at $x_2$.
The boundary conditions on the fields are given by:
\begin{equation}
	\begin{gathered}
{\tilde \phi}_{a}({x_{2,L}^-}) = {\tilde \phi}_{a}({x_{2,L}^+}), \, \, a = 1, \ldots, 8\\
{ \phi}_{a}({x_{2,R}^-}) = { \phi}_{a}({x_{2,R}^+}), \, \, a = 1, \ldots, 8\\
{\tilde \phi}_{9}({x_{2,L}^+}) + {\tilde \phi}_{10}({x_{2,L}^+}) = 2\pi m\\
{\phi}_{9}({x_{2,R}^-})
+ { \phi}_{10}({x_{2,R}^-}) = 2\pi n.
	\end{gathered}
	\label{}
\end{equation}
Here $x^\pm\equiv x\pm \epsilon$ with $\epsilon\rightarrow 0$.

 For a narrow domain wall with
$x_{2} = {x_{2,L}^+}\approx {x_{2,R}^-}$, we can ignore the variation of the fields within the domain wall to find:
\begin{equation}
\label{eqn:boundary-condition}
{\tilde \phi}_{a}({x_2^-}) = \sum_{b=1}^8 (W^8)^{-1}_{ab} \phi_{b}({x_2^+}) - \frac{1}{2} \phi_{1}({x_2^+})
- \pi {m_1} - \pi {n_1} .
\end{equation}
for $a=1,2, \ldots 8$ where $m_1$, $n_1$ are defined in Eq. (\ref{eqn:cos-values}).
As a result of the coefficient of $\frac{1}{2}$ in front of the second term on the right-hand side
of (\ref{eqn:boundary-condition}), any correlation function of the form
$\bigl\langle e^{i \sum_{a=1}^8 n_a {\tilde \phi}_{a}(x<{x_2})}\,  e^{i\sum_{b=1}^8 m_b {\phi}_{b}(x>{x_2})}\bigr\rangle$ vanishes for all integers $m_a$ and all integers $n_a$ satisfying $\sum_{a=1}^8 n_a \equiv 1 \pmod{2}$ (i.e. with odd fermion parity).
Suppose we were to add an arbitrary term of the form
$e^{i \sum_{a=1}^8 p_a {\tilde \phi}_{a}}$ or $e^{i \sum_{a=1}^8 p_a {\phi}_{a}}$ to the action, either
at the domain wall or in the gapped regions to either side of the domain wall. Such a term
could be accounted for in perturbation theory by inserting copies of this term into the correlation function,
but the correlation function will clearly still vanish since such terms cannot cancel the $\frac{1}{2} \phi_{1}({x_2^+})$. Thus,
the `elastic' $S$-matrix vanishes in all odd fermion number sectors. The only way to get a non-vanishing correlation
function is to act with a fermion operator at the domain wall, e.g. 
$\bigl\langle e^{i \sum_{a=1}^8 n_a {\tilde \phi}_{a}(x<{x_2})}\,  e^{i \sum_{b=1}^8 m_b {\phi}_{b}(x>{x_2})}\,
e^{\pm i {\tilde \phi}_{9}({x_2})} \bigr\rangle \neq 0$. Therefore, when a fermionic excitation created in the $I_8$ region passes through the domain wall, its fermion parity gets absorbed by the Majorana zero mode and the electric charge continues into the bosonic $E_8$ region.

\section{Degeneracy splittings}
\label{sec:splittings}

In this section we consider processes that can split the degeneracy that we found in previous sections.
The worst-case scenario would be a splitting that is independent of system size, which would
mean that the degeneracy that we found in Section \ref{sec:degeneracy} is not really stable
against perturbations. However, even if the degeneracy is stable and exact in the thermodynamic limit,
there may be a small finite-size splitting. In this section, we analyze perturbations to the effective action
in Eqs. (\ref{eqn:S_0}) and (\ref{eqn:two-cosines}) in order to determine which ones can split the degeneracy
and how the resulting splitting depends on the system size.

Since it depends on the conservation of fermion parity, any process that changes
 fermion parity in an $I_8$ region will cause
a transition between different ground states. One such process is fermion tunneling from
one $I_8$ region to another. Since all fermions are gapped in the $E_8$ regions, such a process
will cause an exponential splitting $e^{-L/\xi_8}$ where $L$ is the length of the $E_8$ region and
$\xi_8$ is inversely proportional to the gap to $\phi_9$ and $\phi_{10}$ excitations in the $E_8$ regions.

There are also local perturbations in the $I_8$ regions that do not commute with $B_j$ operators, a simple example of which is $\cos(\tilde{\phi}_I+\tilde{\phi}_9)$ where $I=1,2,\cdots,8$. 
Such a perturbation is gapped on either side of the domain wall, but might be present at the interface. 
One might worry that these perturbations completely lift the degeneracy. We postpone the discussion of this issue to Sec \ref{sec:other-gap-terms}. 

\begin{figure}[t!]
	\centering
	\includegraphics[width=0.9\columnwidth]{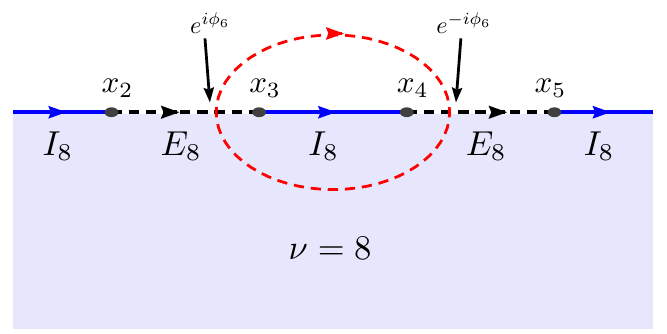}
	\caption{A vortex tunneling process that would cause an energy splitting in a non-chiral system such
as a superconducting nanowire. It does \textit{not} cause such a splitting of the topological degeneracy
associated with ${I_8}-{E_8}$ domain walls at the $\nu=8$ edge. As a result of the chirality of the system,
this process does not measure the fermion parity of the $I_8$ region, as explained in Section \ref{sec:splittings}.}
	\label{fig:vortextul}
\end{figure}

Another fear is that perturbations acting in the $E_8$ regions could cause an energy splitting between states of
different fermion parities by effectively measuring the fermion parity of neighboring $I_8$ regions. Consider, for instance the following charge-neutral perturbation:
\begin{equation}
\label{eqn:ps-pert}
S_{\text{ps}} =  \!\int_{C_{E_8}}\! dx\, dt \,\lambda(x) \,\cos(\phi_6)
\end{equation}
The coupling $\lambda(x)$ could be uniform in $C_{E_8}$; it could be a sum of $\delta$-functions
acting at isolated points (e.g. due to a small number of impurities); or it could be a random function,
due to a random distribution of impurities. All of these possibilites are interesting because
the commutator between $\phi_6$ and the $I_8$ gapping term is an odd multiple of $2\pi i$:
\begin{equation}
	[\phi_6(x), (-\phi_1+\phi_9+3\phi_{10})(x')]=-6\pi i, x<x'.
	\label{eqn:shift-of-cosine}
\end{equation}
Hence, this perturbation can shift the
$u_I$ cosine from one minimum to another. Equivalently, $\cos{\phi_6}(x)$
doesn't commute with the operator $B_1$ if $x_{2,L} < x < x_{3,R}$.
Hence, it can, in principle, give opposite-sign contributions
in second-order perturbation theory to the energies of states with different fermion parities,
i.e. to states with different $A_{2}$ eigenvalues.
The physical picture would be that $S_{\text ps}$ can tunnel a vortex across $C_2$. When it acts again to tunnel a
vortex across $C_4$, it effectively causes a vortex to encircle $C_3$. The state of the system
would then acquire a $\pm 1$, depending on the fermion parity of $C_3$, as occurs in the case
of a superconducting nanowire \cite{Fidkowski11b}.
An alternate description for perturbation theory in $\lambda$ is that it is an instanton gas expansion for
instantons that cause the $I_8$ gap term $\cos(-\phi_1+\phi_9+3\phi_{10})$ to tunnel from one minimum to another.
All of the $\phi_a$ with $a$ even share the property exhibited by $\phi_6$ that $\left(K^{E_8}\right)^{-1}_{a1}$ is odd, leading to a commutation relation similar to (\ref{eqn:shift-of-cosine}).
Hence, any operator $\cos({m_I}{\phi^I})$ in the $E_8$ regions could split states of different $I_8$ fermion parity
so long as $\sum_{I=1}^4 m_{2I}$ is odd. All such perturbations can be analyzed along similar lines,
so we focus on Eq. (\ref{eqn:ps-pert}) for the sake of concreteness. It happens to have the lowest possible
scaling dimension for such a perturbation.

The perturbation shifts the energies of the two ground states.
The energy splitting is the difference in the energy shifts of the two ground states. Consider the
perturbative expansion in powers of $\lambda$. In time-dependent perturbation theory,
the leading-order contribution
to the ground state energy shift is equal to the ground state-to-ground state transition amplitude:
\begin{multline}
\label{eqn:energy-splitting}
\Delta E_{A_2} = -i\int_{-\infty}^\infty dt \int dx \, dx' \, \lambda(x)\, \lambda(x') \\ \times\,
\langle 0_{A_2}| \mathcal{T}(\cos(\phi_{6}(x,t)) \cos(\phi_{6}(x',0)) |0_{A_2}\rangle \\
= \sum_{n\neq 0}  \frac{\left|\langle n| \displaystyle\int dx \, \lambda(x)\cos(\phi_{6}(x,0)) |0_{A_2}\rangle\right|^2}{E_0 - E_n} 
\end{multline}
Here $\mathcal{T}$ represents time-ordering and $n\neq 0$ means summing over all excited states. The subscript ${A_2} =\pm 1$ labels ground states
by their $A_2$ eigenvalues, where
the operator $A_2$ is defined as in Eq. (\ref{eqn:degeneracy-ops}). There will be a non-zero contribution
to the energy splitting when $x_{2,L} < x < x_{3,R}$ and $x_{4,L} < x' < x_{5,R}$ or vice versa,
as in Fig.~\ref{fig:vortextul}.
The second equality follows by a standard spectral decomposition;
the resulting expression is the second-order energy shift in time-independent perturbation theory.

To compute the desired correlation function across the intervening $I_8$ region, we
use the boundary condition (essentially Eq. (\ref{eqn:boundary-condition}), but inverted to
give $\phi_I$ in terms of $\tilde{\phi}_I$) to
rewrite the field $\phi_6$ in terms of the $\tilde{\phi}$ fields and the constants $m_2$ and $n_1$:
\begin{multline}
\phi_{6}({x_3^-}) =  3\pi( {m_2} +  {n_1}) +
\frac{1}{2} \left[ {\tilde \phi}_{1}+\cdots+\tilde{\phi}_7 -  {\tilde \phi}_{8}\right]_{x_3^+}\\
\label{eqn:gapless}
\end{multline}
This equation can be understood as follows: because $\frac{1}{2}(\phi_9+\phi_{10})$ shifts the argument of the $u_I$ cosine by $2\pi$ (it is a single fermion excitation), the combination $\phi_6-\frac{3}{2}(\phi_9+\phi_{10})$ commutes with the $u_I$ cosine term and should be gapless in the $I_8$ region. It is also gapless in the $E_8$ region (since $\phi_9+\phi_{10}$ is pinned), so $\phi_6-\frac{3}{2}(\phi_9+\phi_{10})$ in fact represents a gapless mode across the whole system. If the fields are taken to be very close to $x_3$, we can use the $W^8$ transformation to rewrite it in terms of the $\tilde{\phi}$ modes, which gives exactly the expression in \eqref{eqn:gapless}. This is essentially how
Eq. (\ref{eqn:boundary-condition}) was obtained in Sec~\ref{sec:scattering}. Using
(\ref{eqn:gapless}), we find
\begin{multline}
\label{eqn:vortex-perturbation-x-dep}
\langle 0_{A_2}| \mathcal{T}( e^{i{\phi_6}({x_{3,L}},t)}  e^{-i{\phi_6}({x_{4,R}},0)})| 0_{A_2}\rangle
\sim \\
\frac{A_2}{[t - ({x_3} - {x_4}) - i \delta \,\text{sgn}(t)]^2}
\end{multline}
The points are taken near the ends $x_{3}, x_{4}$ of the intervening
$I_8$ region $C_3$. The numerator on the right-hand-side
is the fermion parity of $C_3$.  For simplicity, we have taken all of the velocities to be equal
and set them to $1$.
The exponent $2$ on the right-hand-side of Eq. (\ref{eqn:vortex-perturbation-x-dep})
must be an even integer since this
is a two-point correlation function of a bosonic operator.
For this particular choice of operator, we happen to find 
the smallest possible exponent for an operator that detects $C_3$ fermion parity.
In addition, $(K^{-1})_{66}=2$, so the $e^{i\phi_6}$
two-point correlation function decays with precisely the same exponent in $C_{E_8}$ as it does across
an $I_8$ region. Consequently, Eq. (\ref{eqn:vortex-perturbation-x-dep}) holds more generally for ${x_3}\in C_2$
and ${x_4}\in C_4$.

Substituting the correlation function (\ref{eqn:vortex-perturbation-x-dep}) into
the expression for the energy shift in Eq. (\ref{eqn:energy-splitting}), we find
an energy splitting $\delta E\equiv\Delta E_{A_{2}=1}-\Delta E_{A_{2}=-1}$
given by:
\begin{equation}
	\begin{split}
\delta E  &= -4i\int_{C_2}\!\!\! dx \int_{C_4}\!\!\! dx' \,
  \int_{-\infty}^\infty dt \frac{\lambda(x)\, \lambda(x') }{[t - ({x} - {x'}) - i \delta\,  \text{sgn}(t)]^2}\\
&=  8\pi  \int_{C_2}\!\!\! dx \int_{C_4}\!\!\! dx'\, \lambda(x)\, \lambda(x')\, \delta({x} - {x'})\\
&=  0
\end{split}
	\label{}
\end{equation}
Hence, this perturbation does not cause any splitting at all!

It is evident that the vanishing of the splitting clearly follows from the chirality of the system, and the result holds to all orders of the perturbation theory. As a comparison,
if we were to replace the correlation function
in this integral by a non-chiral one (e.g. in a nanowire), we would, instead find:
\begin{multline*}
-4i\int_{C_2}\!\!\! dx \int_{C_4} \!\!\! dx' \,  
 \int_{-\infty}^\infty dt \frac{\lambda(x)\, \lambda(x') }{[t^2 - ({x} - {x'})^2 - i \delta]}\\
= 4\pi  \int_{C_2}\!\!\! dx \int_{C_4}\!\!\! dx'\, \frac{\lambda(x)\, \lambda(x')}{|{x} - {x'}|}
\end{multline*}
This result can be understood more intuitively as follows. As we have explained, the only process that can measure the fermion parity (i.e. $A_j$) is encircling the $I_8$ region by an $hc/2e$ vortex (or a phase slip), which can be viewed complementarily
as the virtual tunneling of a fermion from one end of the $I_8$ region to the other.
However, fermions do not actually `tunnel' between the two ends of an $I_8$ region, the reason being that the gapless fermions can only go from the upstream end to the downstream one; not back.
More importantly, {\it every} fermion emitted by the left end {\it must} be absorbed by the right end.
The domain walls and the $C_{I_8}$ bulk are not weakly-coupled; instead, they are a single system
with a single non-local fermion parity.
To get any splitting, we would need to involve the left-moving mode in some way, e.g. to tunnel a fermion from one
$I_8$ region to another. Such processes will contribute exponential splitting.

 
The final source of splitting is fermions tunneling between a metallic lead or localized states in the bulk
and $C_{I_8}$.
The corresponding terms in the action would be:
\begin{equation}
\label{eqn:external-fermions}
S_\text{F} =  \int \! dt\,\sum_{a=1}^8 {v_a}\left({\Psi^\dagger}({x_0},t) \, e^{i\tilde{\phi}_{a}({x_0},t)}
+ \text{h.c.}\right)
\end{equation}
where ${\Psi^\dagger}({x_0},t)$ creates an electron in the metallic lead/low-energy bulk state.
If the spectral function of $\Psi({x_0})$ is independent of energy at low energy, i.e.,
if there is a constant density of states of fermions, then the lifetime of the state, $\tau$, is given by $1/\tau \sim \sum_{a=1}^8 {v^2_a}$, according to Fermi's golden rule. This leads to an exponential decay of fermion parity over time: 
 $\langle {A_i}(t) {A_i}(0)\rangle \sim e^{-t/\tau}$, 
 where $A_i$ is the fermion parity of the $i^{\rm th}$ $I_8$ interval, as defined in Eq. (\ref{eqn:degeneracy-ops}).

\section{More general gap-opening terms}
\label{sec:other-gap-terms}

In the previous section we noticed that local perturbations such as $\cos(\tilde{\phi}_a+\tilde{\phi}_{10})$, $a=1,2,\dots,8$, in $I_8$ regions do not commute with $B$ operators.  In this section we address this issue in a more general setting. We have chosen a particular form of the cosine terms in Eq. (\ref{eqn:two-cosines}) to open a gap to counter-propagating modes, but these are not the unique ways for the sytem to enter the $I_8$ or $E_8$ phases. For instance, a term such as $\cos(\tilde{\phi}_{1} +\tilde{\phi}_{10})$ will drive the system into the $I_8$ phase. Indeed, an arbitrary linear combination of $\cos(\tilde{\phi}_{a} +\tilde{\phi}_{10})$ terms, with $a=1, 2, \ldots, 9$, will also drive the system into the $I_8$ phase, as will more general terms that are not quadratic in the original fermionic variables. Similarly, there is a family of gap-opening terms that will drive the system into the $E_8$ phase. More generally, one can consider a sum of several cosine terms.

The form of the ground state generating operators given in Eq. (\ref{eqn:degeneracy-ops}) depended explicitly on the precise form of the cosine terms. Including more complicated gapping terms, we are no longer able to explicitly construct the ground state generating operators. However, we can argue that the degeneracy is unchanged by ``adiabatically''
deforming the Hamiltonian from the special Hamiltonian considered in Eq. (\ref{eqn:two-cosines})
to any other one that leaves the $C_{I_8}$ and $C_{E_8}$ regions in the $I_8$ and $E_8$ phases
respectively. Since the edge is gapless, the term ``adiabatic deformation'' means a deformation
that does not close the gap to counter-propagating modes.

In order to make this argument, it is useful to distinguish between
two different types of degeneracy that could occur.
By ``topological degeneracy'', we mean states that can only be distinguished by a measurement at two distant points (e.g. two ends of an interval), while ``local degeneracy'' (or ``accidental degeneracy'') will refer to states that can be distinguished by a measurement at a single point. Then a precise statement of our claim is
that the topological degeneracy remains unchanged during any deformation of the system
that does not close the counter-propagating gap. The validity of this claim follows by generalizing
the explicit example in Section \ref{sec:splittings}, which demonstrated how the chirality of the system
protects the degeneracy of states that can only be distinguished by measurements at the two ends of
an interval. Consequently, topological degeneracy can only be lifted when the gap to counter-propagating modes closes (apart from an exponential-in-length splitting due to virtual excitations above the gap to counter-propagating modes). 
Meanwhile, continuously deforming the gap-opening terms (without closing the gap to counter-propagating modes) may cause additional ``local degeneracy'' to develop or be lifted.

We therefore expect that the ground state generating operators also evolve with this adiabatic continuation, while preserving the algebra responsible for the degeneracy.  Consequently the topological degeneracy does not depend on the particular gap-opening terms that are present in the effective action as long as they lead to the desired edge phases. However the construction of the operators in Eq. (\ref{eqn:degeneracy-ops}) is most transparent for particular effective actions with only one cosine gapping term inside each domain, such as Eq. (\ref{eqn:two-cosines}).

\section{Splitting of Superselection Sectors in a Periodic System}
\label{sec:total-droplet}

While we have shown that the splitting between the parity states of a given $I_8$ section is exponentially small in the length of the section, the same is not true of the overall fermion parity of the edge of a quantum Hall droplet, even one with alternating $E_8$ and $I_8$ regions. Note, first, that the action (\ref{eqn:two-cosines}) conserves electrical
charge, so the entire droplet is characterized by its electrical charge, not merely its parity. Therefore, for this
model, the a more precise statement of the question is: how does the splitting between states
of $N$ and $N+1$ electrons depend on the
circumference $L$ of a droplet? On the other hand, the phenomenon described here does not depend on charge
conservation, so we are free to consider models that do not conserve charge. For such a model, we could ask
about the splitting between even and odd fermion parities.

We begin by again considering Eq. \eqref{eqn:boundary-condition}, in which the fermions on the left hand side of the equation (and the domain wall) are related to the bosonic fields on the right. We have already noted that a fermion on one side of the domain wall is completely uncorrelated with any local bosonic operator that one might write down on the other side. What, then, is the fate of a fermionic operator as it evolves along the chiral edge toward the $E_8$ region? The answer may be read off from a careful grouping of the r.h.s. of Eq. \eqref{eqn:boundary-condition} , where we replace $x_{2+}$ by an arbitrary location $x$ within the $E_8$ region. We define the operator
\begin{eqnarray}
\Phi_a(x)&=&\sum_{b=1}^8 (W^8)^{-1}_{ab}\phi_b(x)-\frac12 \phi_1(x)-\pi m_1-\pi n_1\nonumber\\
&=&\sum_{b=1}^{10} (W^8)^{-1}_{ab}\phi_b(x)+\int_{x_2}^x\!\!\mathrm{d}x'\partial_{x'}\!\bigl(\phi_9+2\phi_{10}-\frac12 \phi_1\bigr)\nonumber\\
\end{eqnarray}
It is clear that $\Phi_a(x_{2}^+)=\tilde{\phi}_a(x_{2}^-)$, so that a fermionic operator $e^{i\tilde{\phi}_a}$ becomes $e^{i\Phi_a}$ upon crossing the domain boundary.
Assuming for simplicity that the bosonic fields in the $E_8$ region all have the same velocity (i.e. the velocity matrix is proportional to $I$ in the $E_8$ basis), we may expect that $x$ in the above expression will simply increase with time due to the chiral edge.
From this rearrangement, we can see that despite the appearance of a fraction in front of $\phi_1$, the fermionic fields do, in fact, evolve into allowed operators upon hitting the domain wall into the $E_8$ region. Instead of being a sum of local operators, however, the fermionic fields evolve into a combination of the bosonic fields along with a \emph{non-local} charge measurement operator (the integral in the above expression).  Importantly, while $e^{i\Phi_a}$ has no non-zero correlations with any bosonic operators in the $E_8$ region, it will continue to evolve through that region until it hits another $I_8$ region. Once it does, it will again be allowed to have non-zero correlations with local operators in that region.

However, there is an important difference between a fermion operator that travels from $I_8$ region 1 to $I_8$ region 2 through an $E_8$ region and a fermion operator native to region 2. This involves the final value of the integral term once the operator has traveled through the entire $E_8$ region. We have already seen that at the first domain wall $\phi_9+2\phi_{10}-\frac12 \phi_1$ is pinned to $\pi(m_1+n_1)$. At the second domain wall the pinned value of $n$ for the $E_8$ region must be the same. However, the operator is entering a new $I_8$ region, which may have a new value of $m$. The total value of the integral is therefore $\pi(m_2-m_1)$, and the operator that emerges into the second $I_8$ region is $e^{i\tilde{\phi}_a+i\pi(m_2-m_1)}$. Note that this operator is consistent with conservation of fermion parity within each $I_8$ region. Although $e^{i\tilde{\phi}_a}$ creates a fermion in region 2, $e^{i\pi(m_2-m_1)}$ transfers a fermion from region 2 to region 1, so that the operator as a whole always places the fermion in region 1 despite its evolution.

We now arrive at the source of power law splitting in systems with periodic boundary conditions. If we consider the edge of a quantum Hall droplet with a single $E_8$ and a single $I_8$ region along the edge, then the value of $m_2-m_1$ defined above is necessarily zero, so $e^{i\pi(m_2-m_1)}$ becomes trivial. That is, the electron propagates coherently through the $E_8$ region. The energy cost of adding an electron to the edge of this system therefore has the same dependence on the length of the edge as if the entire edge were in the $I_8$ phase, i.e. it decays as a power law in the circumference. This fact remains true independent of the number of alternating $E_8$ and $I_8$ regions around the edge. While $x$ is in $I_8$ regions, the fermion operator will have the form $e^{i\tilde{\phi}_a(x)+i\pi(m_j-m_0)}$, where the fermion originated in $I_8$ region $0$ and $x$ is currently within $I_8$ region $j$. Once $x$ returns to the original $I_8$ region, the fermion operator becomes local once again. 

Note that this effect splits only states of different \emph{total} fermion parity of the edge, corresponding to the superselection sector of the set of zero modes formed by the domain walls. There is still an exponentially large, exponentially protected Hilbert space due to the presence of the alternating $E_8$ and $I_8$ regions, whose size goes as $2^{\frac{N}2-1}$ in the number $N$ of domain walls.

\section{Measurement via interferometry}
\label{sec:interferometry}

\begin{figure}[t]
 \includegraphics[width=8cm]{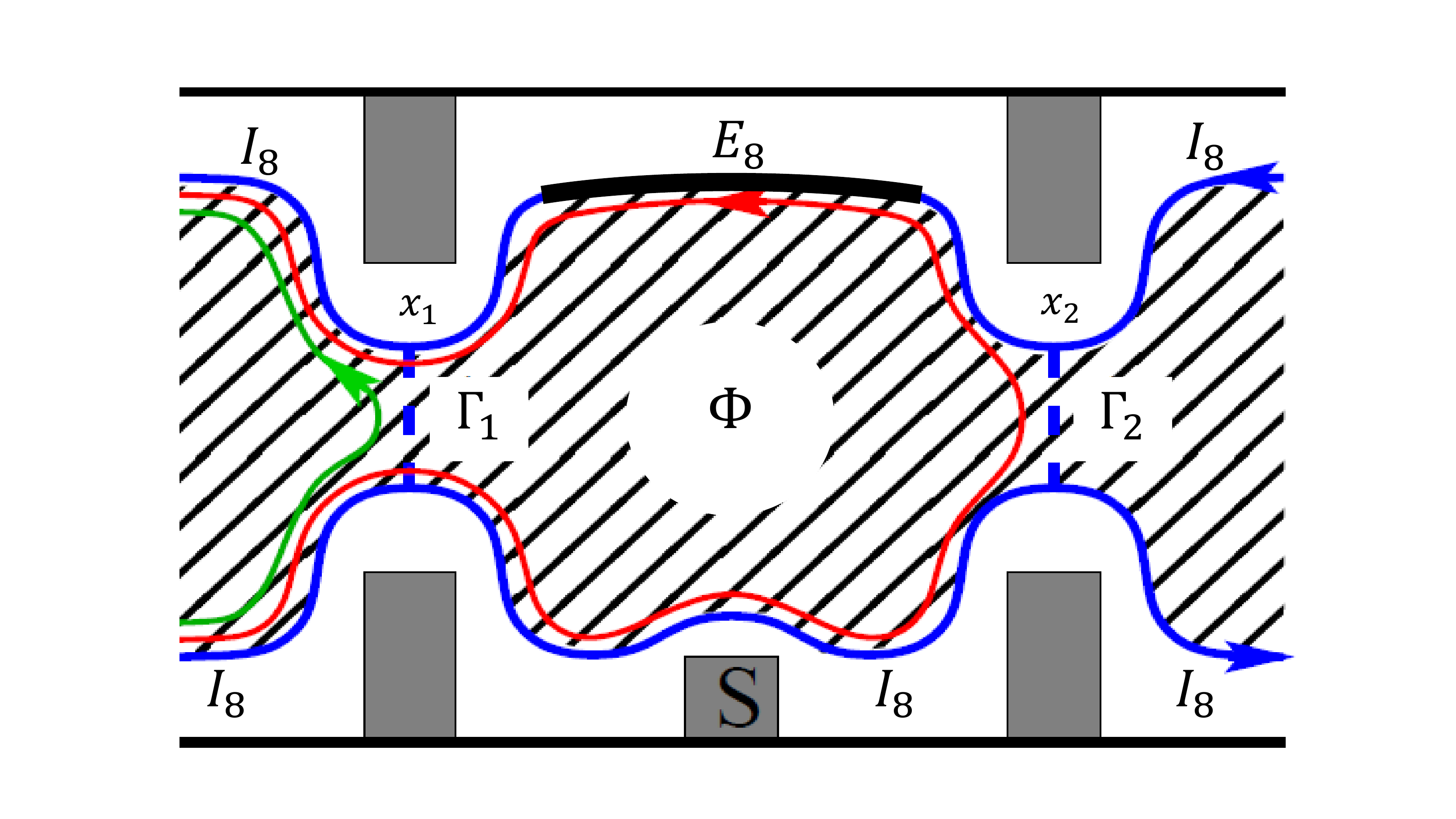}
\label{Fig:E8Interferometer}
\caption{Quantum Hall interferometer reading out the dual fermion parity of an $E_8$ region. The backscattered
current oscillates as a function of flux $\Phi$ through the interferometer, which can be varied with a sidegate $S$
that changes the area of the interference loop. If the interferometer contains a single $E_8$ section (black) and
$I_8$ elsewhere (blue), then the interference pattern undergoes a $\pi$ phase shift if the dual parity
of the $E_8$ region is flipped (i.e. the eigenvalue of the corrresponding $B_j$ operator).
If the interferometer instead contains a single $I_8$ section (black) and $E_8$ elsewhere (blue),
then it similarly measures the fermion parity of the $I_8$ region.}
\end{figure}

It is possible to harness the effect described in the previous section for the purpose of reading out the combined states of qubits formed by collections of $E_8-I_8$ domain walls. To demonstrate, we place an $E_8$ region along one leg of a quantum Hall interferometer, as shown in Fig.~\ref{Fig:E8Interferometer}. We label the pinned values of the surrounding $I_8$ regions by $m_1$ and $m_2$, and the pinned value in the other leg by $m_0$. We assume that there are more $E_8$ regions elsewhere along the edge of the quantum Hall droplet, so that the values of $m_0$, $m_1$, and $m_2$ are unconstrained. We place point contacts at $x_1$ and $x_2$ along the top edge. Further, we assume that the bottom and top edges are connected at a point far to the right and that the edge has total length $L$, so that the bottom contacts are located at points $L-x_1$ and $L-x_2$. 
Our point contact Hamiltonian is therefore 
\begin{equation}
	H_\Gamma=\sum_{j=1}^2\sum_{a,b}\Gamma_{jab} e^{i\tilde{\phi}_a(x_j)+i\pi(m_j-m_0)}e^{-i\tilde{\phi}_b(L-x_j)}+\mathrm{h.c.}
\end{equation} 
where $\Gamma_{jab}$ is the coupling between fermion species $a$ and $b$ at point contact $j$. We can see from this Hamiltonian that the gauge invariant phase difference $\pi(m_2-m_1)$ will appear in any measurement of the current through the interferometer. That is, the state of the qubit stored in the domain walls at the ends of the $E_8$ region may be read out as a $\pi$ phase shift in the interferometric measurement. 

One may similarly read out the fermion parity stored in an $I_8$ region by inverting this setup, replacing each $E_8$ with an $I_8$ and vice-versa. In this case (and in the basis we use), only half of the bosonic modes (i.e. $\phi_{2,4,6,8}$) contribute to the measurement  that could measure the fermion parity in Sec.~\ref{sec:splittings}.

\section{Fractional quantum Hall states: the example of the $\nu=1/8$ bosonic state}
\label{sec:fqh}

There are several avenues along which we can generalize the preceeding construction.
Perhaps the most straightforward would be other fermionic Abelian topological phases (including
fractional ones) that have bosonic edge phases. Such phases are discussed in Ref. \onlinecite{Cano13b}.
Intuitively, domain walls between bosonic and fermionic chiral edge phases of a fermionic bulk state
will support Majorana zero modes, as in the $I_8$-$E_8$ example discussed above, because a
chiral fermion in the fermionic region cannot propagate into the bosonic region and must, therefore,
be absorbed by the domain wall. This can be generalized even further to bulk Abelian topological
phases that have two possible edge phases such that the set of quasiparticles that are gapless in the `smaller' phase is a proper subset of those gapless in the other phase.
Then a domain wall between the two phases must be able to absorb
the quasiparticles that are `missing' from the `smaller' phase.

The simplest example of this is the $K=8$ bosonic state:
\begin{equation}
\label{eqn:K=8=state}
S_{LL} = \int\! dx\, dt \biggl(\frac{8}{4\pi} \partial_t \phi \partial_x \phi -
\frac{1}{4\pi}v\partial_x \phi \partial_x \phi
\biggr).
\end{equation}
This chiral edge has another phase that has only two primary fields. It can be accessed by considering
the enlarged theory:
\begin{equation}
\label{eqn:K=8-stable}
S_{LL} = \int\! dx\, dt \biggl(\frac{1}{4\pi}K_{IJ} \partial_t \phi^I \partial_x \phi^J -
\frac{1}{4\pi}V_{IJ}\partial_x \phi^I \partial_x \phi^J
\biggr).
\end{equation}
where $I,J = 1,2,3$ and
\begin{equation}
K = \begin{pmatrix} 8 & 0 & 0\\ 0 & 0 &1\\ 0 & 1 & 0\end{pmatrix}
\end{equation}
Now consider the following gap terms:
\begin{multline}
\label{eqn:K=8-semion}
S_\text{g} =  \int\! dx\,dt\,  \big[{u_8}\cos \phi_2 +
{u_2}  \cos(8{\phi_1} - 2\phi_{2} + 2\phi_{3})\bigr]
\end{multline}
If the $u_8$ term dominates, then the system is in the $K=8$ phase, as in Eq. (\ref{eqn:K=8=state}).
Suppose, instead, that the $u_2$ term dominates. The remaining gapless excitations are created by
operators $\exp(im^T\phi)$ such that
$m^T\phi$ commutes with the argument of the $u_2$ cosine.
Such operators necessarily have $m_1$ even and, therefore, are semions or multiples of semions. Since the remaining
theory is a fully chiral $c=1$ theory in which the minimal particle is a semion, it must be the $K=2$ theory.
Thus, one edge phase has $\theta=\pi/8$ particles while the other phase has only even numbers of such particles;
the domain walls between such edge phases therefore support a $\mathbb{Z}_2$
topological degeneracy associated with $\pi/8$-particle parity, generalizing the fermion parity of the $I_8$-$E_8$ case.
A $\theta=\pi/8$ particle flowing downstream on the $K=8$ edge will be absorbed by the domain wall since it
cannot be transmitted into the $K=2$ region.
As in the $I_{8}-E_{8}$ case, the splitting will be exponential in the relevant length scales
since chiral perturbations, such as $\cos(\phi_{2} + \phi_{3})$ cannot cause any splitting,
by the argument given in Section \ref{sec:splittings}.

This may be generalized to the $K=2N^2$ theories, which have a $K=2$ edge phase. The domain walls
between such phases support $\mathbb{Z}_N$ zero modes.
\begin{equation}
	S_\mathrm{g}=\!\int dx dt\,[u_{1}(x)\cos\phi_2+u_{2}(x)\cos \left( N(2N\phi_1-\phi_2+\phi_3)\right)].
	\label{eqn:ZN-case-gap-terms}
\end{equation}
where $u_{1}(x)$ and $u_{2}(x)$ are non-zero in alternating intervals,
$u_{1}(x)=u\chi_{o}(x)$, $u_{2}(x)=u\chi_{e}(x)$, analogous to $u_{I}(x), u_{E}(x)$ in
Section \ref{sec:degeneracy}. The degeneracy corresponds to the number modulo $N$ of $\pi/2N^2$
particles that are present in the odd regions.
The ground states must represent the algebra of the operators
\begin{eqnarray}
\label{eqn:degeneracy-ops-frac}
A_j &=& \exp\left[{\frac{i}{N}\int_{x_{2j-1,L}}^{x_{2j,R}} {\partial_x}(2{N^2}\phi_1-N\phi_2+N\phi_3)}\right]\, , \cr
B_j &=& \exp\left[{\frac{i}{N}\int_{x_{2j,L}}^{x_{2j+1,R}} {\partial_x}(\phi_{2})}\right]\\
\end{eqnarray}
which satisfy $A_j ^N = B_j^N = 1$, $A_j  B_{j\pm 1} = e^{\pm 2\pi i/N} B_{j\pm 1}  A_j$.
This algebra implies an $N$-fold degeneracy associated with a pair of domain walls, generalizing the
$2$-fold degeneracy of Eq. (\ref{eqn:K=8-semion}). A two point contact interferometer
can be used to measure the eigenvalues of $A_j$ and $B_j$ in a manner analogous to
that explained in Section \ref{sec:interferometry}.

There is one important subtlety that we did not face in the $N=2$ case.
The operator $B_j$ does not commute with operators such as $\cos \phi_3$ at the domain wall.
However, such an operator opens a gap to the counter-propagating modes in any gapless region.
Hence, if it is present, we can view it as follows.  The size of the domain wall is shrunk due to this
additional term. Meanwhile, the $K=2N^2$ region is expanded and has a spatially-varying gap-opening
term: $\cos\phi_2$ in most of the region and $\cos\phi_3$ in a small part that has been reclaimed from
the domain wall. There is no domain wall between these two parts of the $K=2N^2$ region for the
reasons described in Section \ref{sec:other-gap-terms}.

For an alternative perspective on the innocuousness of
a $\cos\phi_3$ perturbation, note that it would cause an energy splitting between different numbers modulo $N$ of $\pi/2N^2$
particles in the odd regions (i.e. an energy splitting between $A_j$ eigenstates with different eigenvalues).
The same effect can result from a perturbation $\cos({\phi_2}+{\phi_3})$ acting in the even regions
flanking the odd region under consideration. However, such a chiral perturbation causes no splitting
by the arguments in Section \ref{sec:splittings}.

Further generalizations will be discussed elsewhere \cite{Conrad14}.

\section{Domain walls between regions of the same edge phase}
\label{sec:domain-walls-same-qps}

We have seen two nontrivial examples of domain walls carrying topological degeneracy on the edge of an Abelian quantum Hall state. Both examples share a common feature: the edge theory is fully chiral and the particle content of the gapless edge modes do not match on the two sides of the domain wall. Such chirality-protected zero modes occur in Abelian states which contain at least one topologically non-trivial bosonic quasiparticle and in fermionic systems which admit both bosonic and fermionic edge phases. In both of these cases, we can identify the topological degeneracy directly from the mismatch of the quasiparticles in the edge theory.

One might wonder what happens at domain walls between regions of the edge that are described by the same edge phase, but with distinct gapping terms.
An example is the $K^{E_8}\oplus \sigma_z$ edge theory with the gapping terms $u_1(x)\cos(\phi_9+\phi_{10})+u_2(x)\cos(\phi_9-\phi_{10})$. 
This theory is exactly a bosonic $E_8$ edge decoupled from a one-dimensional system of spinless fermions with $p$-wave pairing, which hosts localized Majorana zero modes at the domain walls between superconducting and insulating regions.
We now argue that this is the most generic situation when no other symmetries are present. Since the bulk is short-ranged entangled, there are no fractionalized excitations. The only way to protect zero modes on the edge is through the conservation of fermion parity.
We now consider the $K=K^{E_8}\oplus \sigma_z$ edge with the most general gapping term:
\begin{equation}
	u\cos(\phi_9+\phi_{10})+u'\cos\left(n_I\phi_I\right).
	\label{}
\end{equation}
Let us assume $n$ is a primitive vector. Since we are interested in the case where $\cos( n_I\phi_I)$ gaps the edge to the $E_8$ phase, the criteria in Appendix \ref{sec:find-Keff} requires that $n_9$ and $n_{10}$ must be odd and $n_{1,\dots,8}$ are all even. We construct the following ground-state generating operators:
\begin{equation}
	\begin{gathered}
		A_j = \exp\left[\frac{i}{2}\int_{x_{2j-1,L}}^{x_{2j,R}} {\partial_x}(\phi_{9}+\phi_{10})\right] \\
B_j = \exp\left[\frac{i}{2}\int_{x_{2j,L}}^{x_{2j+1,R}} n_I\partial_x\phi_I\right].
	\end{gathered}
	\label{}
\end{equation}
$A_j$ and $B_j$ are chosen to satisfy $A_j^2=B_j^2=1$. We can easily check that all fermion-parity-conserving local operators commute with them. Their commutation algebra is
\begin{equation}
	A_jB_j=(-1)^{\frac{n_9-n_{10}}{2}}B_jA_j.
	\label{eq:Majorana-degeneracy}
\end{equation}
If $(n_9-n_{10})/2$ is odd, there is topological degeneracy from Majorana zero modes at the domain walls. Otherwise the two phases should be regarded as the same, and the two gapping terms can be continuously deformed into each other, as described in \ref{sec:other-gap-terms}.

We can similarly calculate the topological degeneracy when the gapping terms are $\cos(\phi_9-\phi_{10})$ and $\cos n_I\phi_I$: it is determined by the parity of $(n_9+n_{10})/2$. Because $n_9$ and $n_{10}$ are both odd, $(n_9\pm n_{10})/2$ have opposite parities. Thus, $\cos n_I\phi_I$ is continuously connected to either $\cos(\phi_9+\phi_{10})$ or $\cos(\phi_9-\phi_{10})$. We conclude that there are only two distinct ways to gap the edge.

For the quantum Hall systems we are considering, it is natural to impose $\mathrm{U}(1)$ charge conservation symmetry: without loss of generality, take the charge vector to be $(2t, 1, 1)$ where $t$ is an integer vector. We now prove that if the gapping terms $\cos(n_I\phi_I)$ and $\cos(\phi_9+\phi_{10})$ both yield the $E_8$ phase and conserve charge, they cannot host the Majorana degeneracy (\ref{eq:Majorana-degeneracy}):
the charge of the gapping term is given by 
\begin{equation}
	Q=2\left(\sum_{a=1}^8n_a(K^{E_8})^{-1}t_a+\frac{n_9-n_{10}}{2}\right).
	\label{}
\end{equation}
Because the first term in the parentheses above is even, $Q=0$ requires $(n_9-n_{10})/2$ to also be even. Hence, the $A$ and $B$ operators commute, and there is no degeneracy.



We now present an example
of exponentially-protected Majorana zero modes at domain walls on a chiral charge-conserving edge without gapless chiral fermions. They have the further virtue
that they are impervious to coupling to low-energy or out-of-equilibrium fermions (except at the domain walls),
unlike in the case of the $E_8-I_8$ edge.
We take the bulk phase to be the $\nu=\frac{1}{2}$ strong-pairing state, in which electrons pair up into charge $2e$ Cooper pairs, which then form a $1/8$ bosonic Laughlin state. This edge theory can be described by $K=(8)\oplus \sigma_z$ with charge vector $t=(2,1,1)$. We label the bosonic modes as $\phi_1,\phi_2$, and $\phi_3$ and consider the gapping term: 
\begin{equation}
	S_\mathrm{g}=\int dx d t\, \big[u\cos(\phi_2+\phi_3)+u'\cos(8\phi_1+\phi_2+3\phi_3)\big].
	\label{eq:K8gapping}
\end{equation}
Both gapping vectors are null and charge-conserving. In regions dominated by $u\cos(\phi_2+\phi_3)$, we are left with the low-energy $K=8$ edge mode. When $u'\cos(8\phi_1+\phi_2+3\phi_3)$ dominates, we perform a basis change $\phi=W \tilde{\phi}$, with
\begin{equation}
	W=
	\begin{pmatrix}
 -3 & 0 & -1 \\
 0 & 1 & 0 \\
 8 & 0 & 3 \\
 \end{pmatrix},
	\label{}
\end{equation}
to rewrite the gapping term as $8\phi_1+\phi_2+3\phi_3=\tilde{\phi}_2+\tilde{\phi}_3$ (and $\phi_2+\phi_3=8\tilde{\phi}_1+\tilde{\phi}_2+3\tilde{\phi}_3$); the $K$ matrix is invariant:
\begin{equation}
	W^T
	\begin{pmatrix}
		8 & 0 & 0\\
		0 & 1 & 0\\
		0 & 0 & -1
	\end{pmatrix}W=\begin{pmatrix}
		8 & 0 & 0\\
		0 & 1 & 0\\
		0 & 0 & -1
	\end{pmatrix}.
	\label{}
\end{equation}
Thus, the low-energy theory is still $K=8$ (in the $\tilde{\phi}$ basis), and there are no gapless fermions in either case. 

We now proceed to analyze the domain walls. The coefficients of $\phi_{2,3}$ in (\ref{eq:K8gapping}) are identical to those of $\phi_{9,10}$ in the $I_8$-$E_8$ case, so the ground-state generating operators take the same form as (\ref{eqn:degeneracy-ops}), with $9,10$ replaced by $2,3$. There are $2^{k-1}$ degenerate ground states associated to $2k$ domain walls. 
Let us now consider the length-dependence of the energy splitting. 
Since fermions are gapped everywhere, splitting due to single fermion tunneling is exponentially suppressed. Chiral perturbations -- the only perturbations available at low energies -- will not give
any splitting, as described in Section \ref{sec:splittings} (although, since all of the terms that are left in the low-energy theory take the form $e^{ik\phi_1}$ or $e^{ik\tilde{\phi}_1}$, and such fermion-parity-conserving terms commute with both $A$ and $B$, we need not have worried about these perturbations in this case.)

Further insight comes from the expression of the Majorana zero mode operator (for clarity we neglect the uninteresting local density fluctuations at the domain wall):
\begin{equation}
	\gamma_j = e^{4i{\phi_1}(x_{j,R})}\,e^{i\phi_{3}(x_{j,R})}
\end{equation}
Notice that $e^{4i\phi_1(x_{j,R})}$ operator creates a charge-$e$ excitation, which corresponds to the nontrivial $\mathbb{Z}_2$ boson in the bulk. The existence of this boson allows the Majorana zero mode to be a charge-neutral fermion.
Such an operator is only present in a topologically ordered state.  In general, we may expect that charge-conserving, exponentially protected Majorana zero modes can emerge on the edge of chiral Abelian quantum Hall states which contain a nontrivial $\mathbb{Z}_2$ boson carrying an odd number of electric charges among the quasiparticles. The $\nu=1/2$ strong pairing state turns out to be the simplest Abelian FQH state with this property.

\section{Classification of domain walls on the edge of an Abelian quantum Hall state}
\label{sec:classification}

In the previous sections, we have seen a number of examples of zero modes
and protected degeneracy at domain walls in gapless fully chiral edges of topological phases.
From these examples, we can distill the following general picture.

 Zero modes and protected topological degeneracy can only occur when one or more edge phases have fewer gapless quasiparticle types than there are bulk (gapped) quasiparticle types. (If we also count the gapped quasiparticle types at the edge, then the edge must have precisely the same quasiparticle types as the bulk, but not not all of them must be gapless.) Though necessary, this is not a sufficient condition; there are two different possible sufficient conditions that we give below.

Let us first consider bosonic states. To state the second of these conditions, it is useful to call the two chiral edge phases $1$ and $2$. Moreover, it is useful to consider a configuration with at least $4$ different alternating regions of phases $1$ and $2$, separated by domain walls. The two possible sufficient conditions are:

\begin{enumerate}

\item	The two edge phase phases on either side of a domain wall have different sets of gapless particle types.

\item The two edge phases on either side of a domain wall have the same gapless particle types. There is at least one gapped particle type $a$ such that there is an operator $A$ that transfers an $a$ particle between two phase $1$ regions, and an operator $B$ that transfers a $b$ particle between two phase $2$ regions. The $A$ and $B$ operators should commute with the edge Hamiltonian at low energy (i.e. commute with the argument of the dominant cosine gapping term), and by construction they satisfy $AB=R^{ab}_{a\times b}R^{ba}_{a\times b}BA$. Here $R^{ab}R^{ba}\neq \openone$ represents the full braid phase between $a$ and $b$. Physically, we can view $B$ operators as measuring the number of $a$ particles in region $1$ by braiding.

\end{enumerate}

Case 1 generalizes the $I_8$-$E_8$ domain wall and the domain wall between $K=2N^2$ and $K=2$: there is topological degeneracy due to the mismatch in particle types and the chirality.
Case 2 generalizes the topological degeneracy of parafermionic zero modes in Abelian quantum Hall states~\cite{Barkeshli12, Barkeshli13c, Clarke13a, Lindner12, Cheng12}. 


For fermionic systems, however, there is an additional possibility. Physical fermions can ``pair condense'' on the edge, in which case $a$ and $b$ are just the fermions.  This can lead to Majorana zero modes, as those showing up at domain walls discussed at the end of Section \ref{sec:domain-walls-same-qps} and also the domain walls between the topological superconducting phase and the insulating phase of a nanowire. The commutation algebra between $A$ and $B$ operators, however, do not correspond to the full braiding phase. Heuristically, because full braiding between physical fermions is trivial one has to exchange the fermions to count the fermion parity.


\section{Discussion}
We have found two surprises in this paper: neither superconductivity nor even its analogues are necessary
for Majorana zero modes; and, in spite of the presence of gapless charged excitations in this system,
the splitting between the nearly-degenerate ground states of a collection of these Majorana zero modes
decays exponentially as a function of the relevant length scales.

This means that Majorana zero modes might be observable in a system
composed entirely of GaAs or graphene, without any need for a junction
with a second type of material, namely a
superconductor. It is possible that we have simply traded one difficulty -- the problem
of inducing superconductivity in a semiconductor system -- for
another -- the problem of tuning the edge of the system between phases at will. However, it is alternatively
possible that the latter can be accomplished purely through electrostatic gate control of the edge of the system.
Determining the edge phase in realistic experimental devices is, thus, an important problem for future research.

In order to transform the states associated with multiple domain walls, we need to braid these zero modes. This may,
perhaps, be accomplished by creating corner junctions, as discussed in Ref. \onlinecite{Alicea12a} in the context of the edges of 2D topological insulators with proximity-induced superconductivity. Alternatively, a measurement-only
scheme \cite{Bonderson08b} can be used to manipulate quantum information. Some of the necessary measurements
can be performed via interferometry, as described in Section \ref{sec:interferometry}. However, it will
also be necessary to measure the party of a pair of non-consecutive zero modes. The design of such
a measurement is an important problem for the future.

Although we have focused here on the $\nu=8$ integer quantum Hall state, many fermionic quantum Hall states have
edge phases without gapless fermions \cite{Cano13b}. Integer quantum Hall states in this class must have
$\sigma_{xy} = 8n \frac{e^2}{h}$. However, there are also many fractional quantum Hall examples,
such as the $\nu=3+\frac{1}{5}$ state, which is an observed bulk state \cite{Eisenstein02} --
the edge phase that it exhibits in experiments is, at present, unknown. 
Thus, Majorana zero modes can be created at the edge of all of these states as well.

By folding the field-theoretic description of the edge about a domain wall, we can view
the zero modes discussed here as a conformally-invariant boundary condition for
a conformal field theory in which the right-moving fields are governed by the $I_8$ theory
while the left-moving fields are governed by the $E_8$ theory. It may be possible, by such a mapping, to
connect the quantum information contained in these Majorana zero modes to the boundary
entropy of this conformal boundary condition.

\acknowledgements
We thank Brian Conrad, Lukasz Fidkowski, Matthew Fisher, Roman Lutchyn, and Jon Yard for discussions.
C.N. has been partially supported by the DARPA QuEST
program and the AFOSR under grant FA9550-10-1-0524.
J.C. acknowledges the support of the National Science Foundation
Graduate Research Fellowship under Grant No. DGE1144085. D.J.C. acknowledges the
support of LPS-MPO-CMTC and JQI-NSF-PFC.

\appendix

\section{Useful Matrices}
\label{sec:matrices}
In this appendix we define the matrices refered to in Sec. \ref{sec:edgephases}:
\begin{equation}
K^{E_8} = \begin{pmatrix}
	2 & -1 & 0 & 0 & 0 & 0 & 0 & 0  \\
	-1 & 2 & -1 & 0 & 0 & 0 & -1 & 0 \\
	0 & -1 & 2 & -1 & 0 & 0 & 0 & 0 \\
	0 & 0 & -1 & 2 & -1 & 0 & 0 & 0 \\
	0 & 0 & 0 & -1 & 2 & -1 & 0 & 0 \\
	0 & 0 & 0 & 0 & -1 & 2 & 0 & 0 \\
	0 & -1 & 0 & 0 & 0 & 0 & 2 & -1 \\
	0 & 0 & 0 & 0 & 0 & 0 & -1 & 2 \\	
	\end{pmatrix}
\end{equation}

\begin{equation}
W^8 = \begin{pmatrix}
5 & 5 & 5 & 5 & 5 & 5 & 5 & 5 & 8 & 16\\
10 & 10 & 10 & 9 & 9 & 9 & 9 & 9 & 15 & 30\\
8 & 8 & 8 & 8 & 7 & 7 & 7 & 7 & 12 & 24\\
6 & 6 & 6 & 6 & 6 & 5 & 5 & 5 & 9 & 18\\
4 & 4 & 4 & 4 & 4 & 4 & 3 & 3 & 6 & 12\\
2 & 2 & 2 & 2 & 2 & 2 & 2 & 1 & 3 & 6\\
7 & 7 & 6 & 6 & 6 & 6 & 6 & 6 & 10 & 20\\
4 & 3 & 3 & 3 & 3 & 3 & 3 & 3 & 5 & 10\\
-1 & -1 & -1 & -1 & -1 & -1 & -1 & -1 & -3 & -4\\
2 & 2 & 2 & 2 & 2 & 2 & 2 & 2 & 4 & 7
\end{pmatrix}
\end{equation}

\section{Criterion for gapping to $E_8$}
\label{sec:find-Keff}
Consider the action of Eq.~(\ref{eqn:S_0}). A gapping term will take the form $\cos(n_I\phi_I)$, where the $n_I$ satisfy $n_I (K^{E_8}\oplus \sigma_z)^{-1}_{IJ}n_J = 0$. Here we show how to determine the Lagrangian for the remaining low-energy fields, which will be described by Eq.~(\ref{eqn:bosonic-general}) with some $K$-matrix, $K_{\rm eff}$. Because the low-energy theory has eight gapless modes and satisfies $| \det(K_{\rm eff}) |= 1$, there are only two options: $K_{\rm eff}\sim K^{E_8}$ or $K_{\rm eff} \sim \mathbb{I}_8$, where $\sim$ denotes equality up to $\mathrm{GL}(8,\mathbb{Z})$ transformations. We can distinguish these theories by the absence or presence of quasiparticles with fermionic statistics.

Let $2^g$ be the largest power of 2 which divides $n_9$ and $n_{10}$.  We now show that if either $n_9$ or $n_{10}$ has another factor of 2, then $K_{\rm eff} \sim \mathbb{I}_8$. Otherwise, if $2^{g+1}$ is a factor of $n_I$ for all $I \leq 8$ then $K_{\rm eff}\sim K^{E_8}$, while if not, $K_{\rm eff} \sim \mathbb{I}_8$. In the latter case, the greatest common factor of all $n_I$ is $2^g$, so the gapping vector is not primitive unless $g=0$.

A quasiparticle in the theory described by $K_{\rm eff}$ is labelled by an integer vector $m_I$, which satisfies $m_I (K^{E_8}\oplus \sigma_z)^{-1}_{IJ}n_J = 0$. We consider three cases: first, consider the case where $2^{g+1}$ divides $n_9$. Then a valid quasiparticle in the theory described by $K_{\rm eff}$ has $m_{I\leq 8} = 0, m_9 = 2^{-g} n_{10}, m_{10}=2^{-g} n_9$; since $m_9$ is odd and $m_{10}$ is even, this quasiparticle is a fermion. If we had chosen $2^{g+1}$ to divide $n_{10}$ instead of $n_9$, we could have made a similar construction. Thus, whenever $2^{g+1}$ divides $n_9$ or $n_{10}$, $K_{\rm eff} \sim \mathbb{I}_8$. 

Second, consider the case where $2^{g+1}$ does not divide $n_9$ or $n_{10}$, but $2^{g+1}$ divides $(K^{E_8}\oplus \sigma_z)^{-1}_{IJ}n_J$ for all $I \leq 8$. Then for every $m$ which describes a quasiparticle in the $K_{\rm eff}$ theory, $2^{g+1}$ divides $m_9 n_9 - m_{10}n_{10}$, which requires $m_9 = m_{10}  \mod 2$. Hence, every $m$ describes an even quasiparticle and $K_{\rm eff} \sim K^{E_8}$.

Third, consider the remaining case, where $2^{g+1}$ does not divide $n_9$ or $n_{10}$ and for some $I_0$, $2^{g_0} \leq 2^g$ is the largest power of 2 which divides $(K^{E_8}\oplus\sigma_z)^{-1}_{I_0J}n_J$. Then a valid quasiparticle in the $K_{\rm eff}$ theory has $m_{I_0} = 2^{-g_0} n_{10}$, $m_{10} = 2^{-g_0} (K^{E_8})^{-1}_{I_0J}n_J$ and all other $m_I = 0$. Such an $m$ describes a fermion. Hence, $K_{\rm eff} \sim \mathbb{I}_8$ in this case.

\bibliography{topo-phases}

\begin{thebibliography}{37}
\expandafter\ifx\csname natexlab\endcsname\relax\def\natexlab#1{#1}\fi
\expandafter\ifx\csname bibnamefont\endcsname\relax
  \def\bibnamefont#1{#1}\fi
\expandafter\ifx\csname bibfnamefont\endcsname\relax
  \def\bibfnamefont#1{#1}\fi
\expandafter\ifx\csname citenamefont\endcsname\relax
  \def\citenamefont#1{#1}\fi
\expandafter\ifx\csname url\endcsname\relax
  \def\url#1{\texttt{#1}}\fi
\expandafter\ifx\csname urlprefix\endcsname\relax\def\urlprefix{URL }\fi
\providecommand{\bibinfo}[2]{#2}
\providecommand{\eprint}[2][]{\url{#2}}

\bibitem[{\citenamefont{Alicea}(2012)}]{Alicea12a}
\bibinfo{author}{\bibfnamefont{J.}~\bibnamefont{Alicea}},
  \bibinfo{journal}{Rep. Prog. Phys.} \textbf{\bibinfo{volume}{75}},
  \bibinfo{pages}{076501} (\bibinfo{year}{2012}), \eprint{arXiv:1202.1293}.

\bibitem[{\citenamefont{Beenakker}(2013)}]{Beenakker13a}
\bibinfo{author}{\bibfnamefont{C.~W.~J.} \bibnamefont{Beenakker}},
  \bibinfo{journal}{Annu. Rev. Condens. Matter Phys.}
  \textbf{\bibinfo{volume}{4}}, \bibinfo{pages}{113} (\bibinfo{year}{2013}),
  \eprint{arXiv:1112.1950}.

\bibitem[{\citenamefont{Kitaev}(2003)}]{Kitaev97}
\bibinfo{author}{\bibfnamefont{A.~Y.} \bibnamefont{Kitaev}},
  \bibinfo{journal}{Ann. Phys. (N.Y.)} \textbf{\bibinfo{volume}{303}},
  \bibinfo{pages}{2} (\bibinfo{year}{2003}), \bibinfo{note}{quant-ph/9707021}.

\bibitem[{\citenamefont{Freedman}(1998)}]{Freedman98}
\bibinfo{author}{\bibfnamefont{M.~H.} \bibnamefont{Freedman}},
  \bibinfo{journal}{Proc. Natl. Acad. Sci. U.S.A.}
  \textbf{\bibinfo{volume}{95}}, \bibinfo{pages}{98} (\bibinfo{year}{1998}).

\bibitem[{\citenamefont{Nayak et~al.}(2008)\citenamefont{Nayak, Simon, Stern,
  Freedman, and Sarma}}]{Nayak08}
\bibinfo{author}{\bibfnamefont{C.}~\bibnamefont{Nayak}},
  \bibinfo{author}{\bibfnamefont{S.~H.} \bibnamefont{Simon}},
  \bibinfo{author}{\bibfnamefont{A.}~\bibnamefont{Stern}},
  \bibinfo{author}{\bibfnamefont{M.}~\bibnamefont{Freedman}}, \bibnamefont{and}
  \bibinfo{author}{\bibfnamefont{S.~D.} \bibnamefont{Sarma}},
  \bibinfo{journal}{Rev. Mod. Phys.} \textbf{\bibinfo{volume}{80}},
  \bibinfo{pages}{1083} (\bibinfo{year}{2008}).

\bibitem[{\citenamefont{{Jang} et~al.}(2011)\citenamefont{{Jang}, {Ferguson},
  {Vakaryuk}, {Budakian}, {Chung}, {Goldbart}, and {Maeno}}}]{Jang11}
\bibinfo{author}{\bibfnamefont{J.}~\bibnamefont{{Jang}}},
  \bibinfo{author}{\bibfnamefont{D.~G.} \bibnamefont{{Ferguson}}},
  \bibinfo{author}{\bibfnamefont{V.}~\bibnamefont{{Vakaryuk}}},
  \bibinfo{author}{\bibfnamefont{R.}~\bibnamefont{{Budakian}}},
  \bibinfo{author}{\bibfnamefont{S.~B.} \bibnamefont{{Chung}}},
  \bibinfo{author}{\bibfnamefont{P.~M.} \bibnamefont{{Goldbart}}},
  \bibnamefont{and} \bibinfo{author}{\bibfnamefont{Y.}~\bibnamefont{{Maeno}}},
  \bibinfo{journal}{Science} \textbf{\bibinfo{volume}{331}},
  \bibinfo{pages}{186} (\bibinfo{year}{2011}), \eprint{1101.3611}.

\bibitem[{\citenamefont{Mourik et~al.}(2012)\citenamefont{Mourik, Zuo, Frolov,
  Plissard, Bakkers, and Kouwenhoven}}]{Mourik12}
\bibinfo{author}{\bibfnamefont{V.}~\bibnamefont{Mourik}},
  \bibinfo{author}{\bibfnamefont{K.}~\bibnamefont{Zuo}},
  \bibinfo{author}{\bibfnamefont{S.~M.} \bibnamefont{Frolov}},
  \bibinfo{author}{\bibfnamefont{S.~R.} \bibnamefont{Plissard}},
  \bibinfo{author}{\bibfnamefont{E.~P. A.~M.} \bibnamefont{Bakkers}},
  \bibnamefont{and} \bibinfo{author}{\bibfnamefont{L.~P.}
  \bibnamefont{Kouwenhoven}}, \bibinfo{journal}{Science}
  \textbf{\bibinfo{volume}{336}}, \bibinfo{pages}{1003} (\bibinfo{year}{2012}).

\bibitem[{\citenamefont{Rokhinson et~al.}(2012)\citenamefont{Rokhinson, Liu,
  and Furdyna}}]{Rokhinson12}
\bibinfo{author}{\bibfnamefont{L.~P.} \bibnamefont{Rokhinson}},
  \bibinfo{author}{\bibfnamefont{X.}~\bibnamefont{Liu}}, \bibnamefont{and}
  \bibinfo{author}{\bibfnamefont{J.~K.} \bibnamefont{Furdyna}},
  \bibinfo{journal}{Nat. Phys.} \textbf{\bibinfo{volume}{8}},
  \bibinfo{pages}{795} (\bibinfo{year}{2012}), ISSN \bibinfo{issn}{1745-2473},
  \eprint{arXiv:1204.4212}.

\bibitem[{\citenamefont{Deng et~al.}(2012)\citenamefont{Deng, Yu, Huang,
  Larsson, Caroff, and Xu}}]{Deng12}
\bibinfo{author}{\bibfnamefont{M.~T.} \bibnamefont{Deng}},
  \bibinfo{author}{\bibfnamefont{C.~L.} \bibnamefont{Yu}},
  \bibinfo{author}{\bibfnamefont{G.~Y.} \bibnamefont{Huang}},
  \bibinfo{author}{\bibfnamefont{M.}~\bibnamefont{Larsson}},
  \bibinfo{author}{\bibfnamefont{P.}~\bibnamefont{Caroff}}, \bibnamefont{and}
  \bibinfo{author}{\bibfnamefont{H.~Q.} \bibnamefont{Xu}},
  \bibinfo{journal}{Nano Letters} \textbf{\bibinfo{volume}{12}},
  \bibinfo{pages}{6414} (\bibinfo{year}{2012}), \eprint{arXiv:1204.4130}.

\bibitem[{\citenamefont{Churchill et~al.}(2013)\citenamefont{Churchill, Fatemi,
  Grove-Rasmussen, Deng, Caroff, Xu, and Marcus}}]{Churchill13}
\bibinfo{author}{\bibfnamefont{H.~O.~H.} \bibnamefont{Churchill}},
  \bibinfo{author}{\bibfnamefont{V.}~\bibnamefont{Fatemi}},
  \bibinfo{author}{\bibfnamefont{K.}~\bibnamefont{Grove-Rasmussen}},
  \bibinfo{author}{\bibfnamefont{M.~T.} \bibnamefont{Deng}},
  \bibinfo{author}{\bibfnamefont{P.}~\bibnamefont{Caroff}},
  \bibinfo{author}{\bibfnamefont{H.~Q.} \bibnamefont{Xu}}, \bibnamefont{and}
  \bibinfo{author}{\bibfnamefont{C.~M.} \bibnamefont{Marcus}},
  \bibinfo{journal}{Phys. Rev. B} \textbf{\bibinfo{volume}{87}},
  \bibinfo{pages}{241401} (\bibinfo{year}{2013}), \eprint{arXiv:1303.2407}.

\bibitem[{\citenamefont{Das et~al.}(2012)\citenamefont{Das, Ronen, Most, Oreg,
  Heiblum, and Shtrikman}}]{Das12}
\bibinfo{author}{\bibfnamefont{A.}~\bibnamefont{Das}},
  \bibinfo{author}{\bibfnamefont{Y.}~\bibnamefont{Ronen}},
  \bibinfo{author}{\bibfnamefont{Y.}~\bibnamefont{Most}},
  \bibinfo{author}{\bibfnamefont{Y.}~\bibnamefont{Oreg}},
  \bibinfo{author}{\bibfnamefont{M.}~\bibnamefont{Heiblum}}, \bibnamefont{and}
  \bibinfo{author}{\bibfnamefont{H.}~\bibnamefont{Shtrikman}},
  \bibinfo{journal}{Nat. Phys.} \textbf{\bibinfo{volume}{8}},
  \bibinfo{pages}{887} (\bibinfo{year}{2012}), \eprint{arXiv:1205.7073}.

\bibitem[{\citenamefont{Finck et~al.}(2013)\citenamefont{Finck, Van~Harlingen,
  Mohseni, Jung, and Li}}]{Finck12}
\bibinfo{author}{\bibfnamefont{A.~D.~K.} \bibnamefont{Finck}},
  \bibinfo{author}{\bibfnamefont{D.~J.} \bibnamefont{Van~Harlingen}},
  \bibinfo{author}{\bibfnamefont{P.~K.} \bibnamefont{Mohseni}},
  \bibinfo{author}{\bibfnamefont{K.}~\bibnamefont{Jung}}, \bibnamefont{and}
  \bibinfo{author}{\bibfnamefont{X.}~\bibnamefont{Li}}, \bibinfo{journal}{Phys.
  Rev. Lett.} \textbf{\bibinfo{volume}{110}}, \bibinfo{pages}{126406}
  (\bibinfo{year}{2013}).

\bibitem[{\citenamefont{Das~Sarma et~al.}(2006)\citenamefont{Das~Sarma, Nayak,
  and Tewari}}]{DasSarma06a}
\bibinfo{author}{\bibfnamefont{S.}~\bibnamefont{Das~Sarma}},
  \bibinfo{author}{\bibfnamefont{C.}~\bibnamefont{Nayak}}, \bibnamefont{and}
  \bibinfo{author}{\bibfnamefont{S.}~\bibnamefont{Tewari}},
  \bibinfo{journal}{Phys. Rev. B} \textbf{\bibinfo{volume}{73}},
  \bibinfo{pages}{220502} (\bibinfo{year}{2006}).

\bibitem[{\citenamefont{{Fu} and {Kane}}(2008)}]{Fu08}
\bibinfo{author}{\bibfnamefont{L.}~\bibnamefont{{Fu}}} \bibnamefont{and}
  \bibinfo{author}{\bibfnamefont{C.~L.} \bibnamefont{{Kane}}},
  \bibinfo{journal}{Phys. Rev. Lett.} \textbf{\bibinfo{volume}{100}},
  \bibinfo{pages}{096407} (\bibinfo{year}{2008}), \eprint{0707.1692}.

\bibitem[{\citenamefont{{Sau} et~al.}(2010)\citenamefont{{Sau}, {Lutchyn},
  {Tewari}, and {Das Sarma}}}]{Sau10a}
\bibinfo{author}{\bibfnamefont{J.~D.} \bibnamefont{{Sau}}},
  \bibinfo{author}{\bibfnamefont{R.~M.} \bibnamefont{{Lutchyn}}},
  \bibinfo{author}{\bibfnamefont{S.}~\bibnamefont{{Tewari}}}, \bibnamefont{and}
  \bibinfo{author}{\bibfnamefont{S.}~\bibnamefont{{Das Sarma}}},
  \bibinfo{journal}{Phys. Rev. Lett.} \textbf{\bibinfo{volume}{104}},
  \bibinfo{pages}{040502} (\bibinfo{year}{2010}), \eprint{0907.2239}.

\bibitem[{\citenamefont{{Lutchyn} et~al.}(2010)\citenamefont{{Lutchyn}, {Sau},
  and {Das Sarma}}}]{Lutchyn10}
\bibinfo{author}{\bibfnamefont{R.~M.} \bibnamefont{{Lutchyn}}},
  \bibinfo{author}{\bibfnamefont{J.~D.} \bibnamefont{{Sau}}}, \bibnamefont{and}
  \bibinfo{author}{\bibfnamefont{S.}~\bibnamefont{{Das Sarma}}},
  \bibinfo{journal}{Phys. Rev. Lett.} \textbf{\bibinfo{volume}{105}},
  \bibinfo{pages}{077001} (\bibinfo{year}{2010}), \eprint{1002.4033}.

\bibitem[{\citenamefont{{Oreg} et~al.}(2010)\citenamefont{{Oreg}, {Refael}, and
  {von Oppen}}}]{Oreg10}
\bibinfo{author}{\bibfnamefont{Y.}~\bibnamefont{{Oreg}}},
  \bibinfo{author}{\bibfnamefont{G.}~\bibnamefont{{Refael}}}, \bibnamefont{and}
  \bibinfo{author}{\bibfnamefont{F.}~\bibnamefont{{von Oppen}}},
  \bibinfo{journal}{Phys. Rev. Lett.} \textbf{\bibinfo{volume}{105}},
  \bibinfo{pages}{177002} (\bibinfo{year}{2010}), \eprint{1003.1145}.

\bibitem[{\citenamefont{{Fidkowski} et~al.}(2011)\citenamefont{{Fidkowski},
  {Lutchyn}, {Nayak}, and {Fisher}}}]{Fidkowski11b}
\bibinfo{author}{\bibfnamefont{L.}~\bibnamefont{{Fidkowski}}},
  \bibinfo{author}{\bibfnamefont{R.~M.} \bibnamefont{{Lutchyn}}},
  \bibinfo{author}{\bibfnamefont{C.}~\bibnamefont{{Nayak}}}, \bibnamefont{and}
  \bibinfo{author}{\bibfnamefont{M.~P.~A.} \bibnamefont{{Fisher}}},
  \bibinfo{journal}{\prb} \textbf{\bibinfo{volume}{84}},
  \bibinfo{pages}{195436} (\bibinfo{year}{2011}).

\bibitem[{\citenamefont{{Sau} et~al.}(2011)\citenamefont{{Sau}, {Halperin},
  {Flensberg}, and {Das Sarma}}}]{Sau11}
\bibinfo{author}{\bibfnamefont{J.~D.} \bibnamefont{{Sau}}},
  \bibinfo{author}{\bibfnamefont{B.~I.} \bibnamefont{{Halperin}}},
  \bibinfo{author}{\bibfnamefont{K.}~\bibnamefont{{Flensberg}}},
  \bibnamefont{and} \bibinfo{author}{\bibfnamefont{S.}~\bibnamefont{{Das
  Sarma}}}, \bibinfo{journal}{Phys. Rev. B} \textbf{\bibinfo{volume}{84}},
  \bibinfo{pages}{144509} (\bibinfo{year}{2011}).

\bibitem[{\citenamefont{Moore and Read}(1991)}]{Moore91}
\bibinfo{author}{\bibfnamefont{G.}~\bibnamefont{Moore}} \bibnamefont{and}
  \bibinfo{author}{\bibfnamefont{N.}~\bibnamefont{Read}},
  \bibinfo{journal}{Nucl. Phys. B} \textbf{\bibinfo{volume}{360}},
  \bibinfo{pages}{362} (\bibinfo{year}{1991}).

\bibitem[{\citenamefont{Greiter et~al.}(1992)\citenamefont{Greiter, Wen, and
  Wilczek}}]{Greiter92}
\bibinfo{author}{\bibfnamefont{M.}~\bibnamefont{Greiter}},
  \bibinfo{author}{\bibfnamefont{X.~G.} \bibnamefont{Wen}}, \bibnamefont{and}
  \bibinfo{author}{\bibfnamefont{F.}~\bibnamefont{Wilczek}},
  \bibinfo{journal}{Nucl. Phys. B} \textbf{\bibinfo{volume}{374}},
  \bibinfo{pages}{567} (\bibinfo{year}{1992}).

\bibitem[{\citenamefont{Nayak and Wilczek}(1996)}]{Nayak96c}
\bibinfo{author}{\bibfnamefont{C.}~\bibnamefont{Nayak}} \bibnamefont{and}
  \bibinfo{author}{\bibfnamefont{F.}~\bibnamefont{Wilczek}},
  \bibinfo{journal}{Nucl. Phys. B} \textbf{\bibinfo{volume}{479}},
  \bibinfo{pages}{529} (\bibinfo{year}{1996}),
  \bibinfo{note}{cond-mat/9605145}.

\bibitem[{\citenamefont{Read and Rezayi}(1996)}]{Read96}
\bibinfo{author}{\bibfnamefont{N.}~\bibnamefont{Read}} \bibnamefont{and}
  \bibinfo{author}{\bibfnamefont{E.}~\bibnamefont{Rezayi}},
  \bibinfo{journal}{Phys. Rev. B} \textbf{\bibinfo{volume}{54}},
  \bibinfo{pages}{16864} (\bibinfo{year}{1996}),
  \bibinfo{note}{cond-mat/9609079}.

\bibitem[{\citenamefont{{Bonderson} et~al.}(2011)\citenamefont{{Bonderson},
  {Gurarie}, and {Nayak}}}]{Bonderson11a}
\bibinfo{author}{\bibfnamefont{P.}~\bibnamefont{{Bonderson}}},
  \bibinfo{author}{\bibfnamefont{V.}~\bibnamefont{{Gurarie}}},
  \bibnamefont{and} \bibinfo{author}{\bibfnamefont{C.}~\bibnamefont{{Nayak}}},
  \bibinfo{journal}{\prb} \textbf{\bibinfo{volume}{83}},
  \bibinfo{pages}{075303} (\bibinfo{year}{2011}).

\bibitem[{\citenamefont{Lee et~al.}(2007)\citenamefont{Lee, Ryu, Nayak, and
  Fisher}}]{LeeSS07}
\bibinfo{author}{\bibfnamefont{S.-S.} \bibnamefont{Lee}},
  \bibinfo{author}{\bibfnamefont{S.}~\bibnamefont{Ryu}},
  \bibinfo{author}{\bibfnamefont{C.}~\bibnamefont{Nayak}}, \bibnamefont{and}
  \bibinfo{author}{\bibfnamefont{M.~P.~A.} \bibnamefont{Fisher}},
  \bibinfo{journal}{Phys. Rev. Lett.} \textbf{\bibinfo{volume}{99}},
  \bibinfo{pages}{236807} (\bibinfo{year}{2007}), \eprint{arXiv:0707.0478}.

\bibitem[{\citenamefont{Levin et~al.}(2007)\citenamefont{Levin, Halperin, and
  Rosenow}}]{Levin07}
\bibinfo{author}{\bibfnamefont{M.}~\bibnamefont{Levin}},
  \bibinfo{author}{\bibfnamefont{B.~I.} \bibnamefont{Halperin}},
  \bibnamefont{and} \bibinfo{author}{\bibfnamefont{B.}~\bibnamefont{Rosenow}},
  \bibinfo{journal}{Phys. Rev. Lett.} \textbf{\bibinfo{volume}{99}},
  \bibinfo{eid}{236806} (\bibinfo{year}{2007}), \eprint{arXiv:0707.0483}.

\bibitem[{\citenamefont{Barkeshli and Qi}(2012)}]{Barkeshli12}
\bibinfo{author}{\bibfnamefont{M.}~\bibnamefont{Barkeshli}} \bibnamefont{and}
  \bibinfo{author}{\bibfnamefont{X.-L.} \bibnamefont{Qi}},
  \bibinfo{journal}{Phys. Rev. X} \textbf{\bibinfo{volume}{2}},
  \bibinfo{pages}{031013} (\bibinfo{year}{2012}), \eprint{arXiv:1112.3311}.

\bibitem[{\citenamefont{Barkeshli
  et~al.}(2013{\natexlab{a}})\citenamefont{Barkeshli, Jian, and
  Qi}}]{Barkeshli13a}
\bibinfo{author}{\bibfnamefont{M.}~\bibnamefont{Barkeshli}},
  \bibinfo{author}{\bibfnamefont{C.-M.} \bibnamefont{Jian}}, \bibnamefont{and}
  \bibinfo{author}{\bibfnamefont{X.-L.} \bibnamefont{Qi}},
  \bibinfo{journal}{Phys. Rev. B} \textbf{\bibinfo{volume}{87}},
  \bibinfo{pages}{045130} (\bibinfo{year}{2013}{\natexlab{a}}),
  \eprint{arXiv:1208.4834}.

\bibitem[{\citenamefont{Bombin}(2010)}]{bombin2010}
\bibinfo{author}{\bibfnamefont{H.}~\bibnamefont{Bombin}},
  \bibinfo{journal}{Phys. Rev. Lett.} \textbf{\bibinfo{volume}{105}},
  \bibinfo{pages}{030403} (\bibinfo{year}{2010}), \eprint{arXiv:1004.1838}.

\bibitem[{\citenamefont{Cano et~al.}(2014)\citenamefont{Cano, Cheng, Mulligan,
  Nayak, Plamadeala, and Yard}}]{Cano13b}
\bibinfo{author}{\bibfnamefont{J.}~\bibnamefont{Cano}},
  \bibinfo{author}{\bibfnamefont{M.}~\bibnamefont{Cheng}},
  \bibinfo{author}{\bibfnamefont{M.}~\bibnamefont{Mulligan}},
  \bibinfo{author}{\bibfnamefont{C.}~\bibnamefont{Nayak}},
  \bibinfo{author}{\bibfnamefont{E.}~\bibnamefont{Plamadeala}},
  \bibnamefont{and} \bibinfo{author}{\bibfnamefont{J.}~\bibnamefont{Yard}},
  \bibinfo{journal}{Phys. Rev. B} \textbf{\bibinfo{volume}{89}},
  \bibinfo{pages}{115116} (\bibinfo{year}{2014}), \eprint{arXiv:1310.5708}.

\bibitem[{\citenamefont{Clarke et~al.}(2013)\citenamefont{Clarke, Alicea, and
  Shtengel}}]{Clarke13a}
\bibinfo{author}{\bibfnamefont{D.~J.} \bibnamefont{Clarke}},
  \bibinfo{author}{\bibfnamefont{J.}~\bibnamefont{Alicea}}, \bibnamefont{and}
  \bibinfo{author}{\bibfnamefont{K.}~\bibnamefont{Shtengel}},
  \bibinfo{journal}{Nat. Commun.} \textbf{\bibinfo{volume}{4}},
  \bibinfo{pages}{1348} (\bibinfo{year}{2013}), \eprint{arXiv:1204.5479}.

\bibitem[{\citenamefont{Conrad et~al.}()\citenamefont{Conrad, Nayak, and
  Yard}}]{Conrad14}
\bibinfo{author}{\bibfnamefont{B.}~\bibnamefont{Conrad}},
  \bibinfo{author}{\bibfnamefont{C.}~\bibnamefont{Nayak}}, \bibnamefont{and}
  \bibinfo{author}{\bibfnamefont{J.}~\bibnamefont{Yard}}, \bibinfo{note}{in
  prep.}

\bibitem[{\citenamefont{Barkeshli
  et~al.}(2013{\natexlab{b}})\citenamefont{Barkeshli, Jian, and
  Qi}}]{Barkeshli13c}
\bibinfo{author}{\bibfnamefont{M.}~\bibnamefont{Barkeshli}},
  \bibinfo{author}{\bibfnamefont{C.-M.} \bibnamefont{Jian}}, \bibnamefont{and}
  \bibinfo{author}{\bibfnamefont{X.-L.} \bibnamefont{Qi}},
  \bibinfo{journal}{Phys. Rev. B} \textbf{\bibinfo{volume}{88}},
  \bibinfo{pages}{235103} (\bibinfo{year}{2013}{\natexlab{b}}).

\bibitem[{\citenamefont{Lindner et~al.}(2012)\citenamefont{Lindner, Berg,
  Refael, and Stern}}]{Lindner12}
\bibinfo{author}{\bibfnamefont{N.~H.} \bibnamefont{Lindner}},
  \bibinfo{author}{\bibfnamefont{E.}~\bibnamefont{Berg}},
  \bibinfo{author}{\bibfnamefont{G.}~\bibnamefont{Refael}}, \bibnamefont{and}
  \bibinfo{author}{\bibfnamefont{A.}~\bibnamefont{Stern}},
  \bibinfo{journal}{Phys. Rev. X} \textbf{\bibinfo{volume}{2}},
  \bibinfo{pages}{041002} (\bibinfo{year}{2012}), \eprint{arXiv:1204.5733}.

\bibitem[{\citenamefont{Cheng}(2012)}]{Cheng12}
\bibinfo{author}{\bibfnamefont{M.}~\bibnamefont{Cheng}},
  \bibinfo{journal}{Phys. Rev. B} \textbf{\bibinfo{volume}{86}},
  \bibinfo{pages}{195126} (\bibinfo{year}{2012}), \eprint{arXiv:1204.6084}.

\bibitem[{\citenamefont{Bonderson et~al.}(2008)\citenamefont{Bonderson,
  Freedman, and Nayak}}]{Bonderson08b}
\bibinfo{author}{\bibfnamefont{P.}~\bibnamefont{Bonderson}},
  \bibinfo{author}{\bibfnamefont{M.}~\bibnamefont{Freedman}}, \bibnamefont{and}
  \bibinfo{author}{\bibfnamefont{C.}~\bibnamefont{Nayak}},
  \bibinfo{journal}{Phys. Rev. Lett.} \textbf{\bibinfo{volume}{101}},
  \bibinfo{pages}{010501} (\bibinfo{year}{2008}), \eprint{arXiv:0802.0279}.

\bibitem[{\citenamefont{Eisenstein et~al.}(2002)\citenamefont{Eisenstein,
  Cooper, Pfeiffer, and West}}]{Eisenstein02}
\bibinfo{author}{\bibfnamefont{J.~P.} \bibnamefont{Eisenstein}},
  \bibinfo{author}{\bibfnamefont{K.~B.} \bibnamefont{Cooper}},
  \bibinfo{author}{\bibfnamefont{L.~N.} \bibnamefont{Pfeiffer}},
  \bibnamefont{and} \bibinfo{author}{\bibfnamefont{K.~W.} \bibnamefont{West}},
  \bibinfo{journal}{Phys. Rev. Lett.} \textbf{\bibinfo{volume}{88}},
  \bibinfo{pages}{076801} (\bibinfo{year}{2002}).

\end{thebibliography}

\end{document}